\documentclass[%
 reprint,
superscriptaddress,
 amsmath,amssymb,
 aps,
]{revtex4-1}

\usepackage{graphicx}
\usepackage{hyperref}
\usepackage[version=4]{mhchem}
\usepackage{float}
\usepackage{caption}
\usepackage{subcaption}
\usepackage{tikz}
\usepackage[outline]{contour}
\contourlength{0.5pt}
\definecolor{magenta}{rgb}{1,0,1}
\hypersetup{
	colorlinks   = true,
	citecolor    = blue,
	linkcolor = blue,
    urlcolor = blue
}

\tikzset{
  double arrow/.style args={#1 colored by #2 and #3}{
    -stealth,line width=#1,#2, 
    postaction={draw,-stealth,#3,line width=(#1)/3,
                shorten <=(#1)/3,shorten >=2*(#1)/3}, 
  }
}

\usepackage{color}

\begin{document}
\title{Two-Dimensional Transition Metal Silicate Formed on Ru (0001) by Hydrogenation}
\author{Kayahan Saritas}
\affiliation{Department of Applied Physics, Yale University, New Haven, CT, 06520}
\author{Eric I. Altman}
\affiliation{Department of Chemistry, Yale University, New Haven, CT, 06520}
\author{Sohrab Ismail-Beigi}
\affiliation{Department of Applied Physics, Yale University, New Haven, CT, 06520}
\email{sohrab.ismail-beigi@yale.edu}

\begin{abstract}
Two-dimensional (2D) transition metal silicates are interesting materials because of their potential ferromagnetism together with their inherent piezoelectric response to their structural symmetry. Substrate-assisted bottom-up synthesis of these materials offers flexibility in their chemical composition.  However, synthesizing free-standing layers has been challenging due to strong overlayer-substrate interactions which hinders exfoliation of the overlayer. 
Here, using density functional theory calculations, we systematically investigate the hydrogenation of such overlayers as a way to decrease the substrate-overlayer interactions. 
Using the Fe$_2$Si$_2$O$_8\cdot$O/Ru(0001) structure as our starting point, we study hydrogenation levels up to Fe$_2$Si$_2$O$_9$H$_4$/Ru(0001). 
Structural and thermodynamic properties are studied at different hydrogenation levels to describe the conditions under which exfoliation of the Fe-silicate is feasible. Simulated surface core-level binding energy shifts of core electrons show that Fe is primarily in the 3+ state throughout the hydrogenation process. 
Simulated reflection adsorption infrared spectroscopy (RAIRS) yields distinctive shifts in vibrational peaks with increasing hydrogenation which can guide future experiments.
\end{abstract}

\maketitle

\section{Introduction}

Two-dimensional (2D) materials research has grown immensely in recent years: a key fact fueling these developments is that these materials can be isolated from their bulk counterparts. 
These 2D materials show distinct physical and chemical properties from their bulk counterparts due to effects such as quantum confinement and significantly increased exposed surface area. 
Top-down approaches, such as mechanical exfoliation, are often used to isolate 2D layers from the bulk. Bottom-up approaches, however, can enable control over the structure and chemical composition of the synthesized layers and hence provide more flexibility. By selecting an appropriate substrate, surface-assisted bottom-up synthesis can create structures that are not available in bulk phases, e.g., 2D bilayer silicates \cite{Buchner2017}. For the large-scale application of these bottom-up fabricated materials, they must be high quality and are air-stable.

Recently, 2D transition metal silicates such as Fe, Ni, and Ti-silicates have been synthesized on metal substrates \cite{Wodarczyk2013, Zhou2019, Fischer2015, Li2017, Doudin2021}. These materials have the chemical formula of M$_2$Si$_2$O$_8$, where a honeycomb Si$_2$O$_5$ layer with corner-sharing tetrahedra is stacked on top of a transition metal oxide layer which has both edge and corner-sharing 5-fold coordinated, square pyramidal polyhedra. 
These transition metal silicates on metal substrates resemble crystal structures of naturally existing sheet silicates (phyllosilicates), particularly that of dehydroxylated nontronite \cite{Wodarczyk2013, Tissot2016}. 
These structures are also very similar despite the nominal preferences of Ni, Fe, and Ti for different oxidation states.
The synthesis procedure for these transition metal silicates starts with depositing Si or SiO and the transition metal at modest temperatures on the substrate or by depositing the Si on an alloy that contains the target transition metal and then following annealing near 950 K \cite{Zhou2019}. 

To our knowledge, there has been no concerted effort to yield a VDW layer from these materials after their syntheses: The 2D transition metal silicate forms strong chemical bonds with the substrate, which prevents the isolation of the 2D layer \cite{Zhou2019}. 
An efficient strategy can be to hydrogenate these materials with a hydrogen source such as hydrogen gas,  plasma, or water. The dehydroxylation of kaolinite to yield metakaolin (M$_2$Si$_2$O$_9$) is done in large-scale industrial processes \cite{White}, hence these hydrogenation reactions can be chemically accessible.

In this work, using density functional theory (DFT), we study the hydrogenation mechanism of 2D Fe-silicates on Ru (0001). 2D Fe-silicate on Ru (0001) \cite{Wodarczyk2013} and Pd (111)\cite{Saritas2021a} were synthesized previously. We show that Fe-silicate/substrate bond can be substantially weakened by the hydrogenation, such that the film may be isolated to yield a 2D Van der Waals (VDW) layer. Most importantly, we show that the interface oxygens, which are the oxygens chemically bound on the substrate surface, behave in a way that is similar to the hydroxide groups during the water adsorption and dissociation on Ru (0001) \cite{Feibelman2002, Michaelides2003, Tatarkhanov}. With increasing hydrogenation, oxygen atoms originally near Ru (0001) hcp and bridge sites migrate closer to atop sites in our materials, and this is similar to hydroxide groups forming on the atop Ru (0001) sites in the case of the water adsorption. Fully hydrogenated Fe-silicate forms a compound with the chemical formula of Fe$_2$Si$_2$O$_9$H$_4$, thus incorporating the adatom oxygen as well which was initially located at the hcp site of Ru (0001). Fe$_2$Si$_2$O$_9$H$_4$ on Ru (0001) has a binding energy of 0.101 meV/{\AA}$^2$, which classifies the compound as potentially exfoliable \cite{Mounet2018}. Our analyses show that the Fe atom has a charge state of nearly $3+$ for all Fe$_2$Si$_2$O$_9$H$_n$ ($0\leq{n}\leq4$) on Ru(0001) complexes we studied. We perform density of states, workfunction, core level binding energy shifts, and simulated reflection adsorption infrared spectroscopy  (RAIRS) calculations plus structural analyses to corroborate our findings with ongoing experiments.

\setlength{\fboxsep}{0pt}%
\setlength{\fboxrule}{2pt}%
\begin{figure*}[ht]
\centering
    \begin{subfigure}[htbp]{0.31\textwidth}
        \begin{tikzpicture}
        \node[inner sep=0pt] 
        {\fbox{\includegraphics[width=\textwidth]{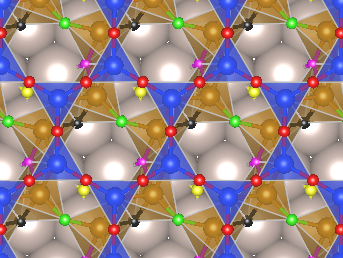}}};
        \node[align=center,fill=white,draw] at (-2.4, 1.8) {(a)};
        \draw[black, ultra thick] (-1.85,-0.57) -- (-0.9,1.05);
        \draw[black, ultra thick] (-1.85,-0.57) -- (0,-0.57);
        \draw[black, ultra thick] (-0.9,1.05) -- (0.95,1.05);
        \draw[black, ultra thick] (0,-0.57) -- (0.95,1.05);
        \end{tikzpicture}
    \end{subfigure}
    \begin{subfigure}[htbp]{0.31\textwidth}
        \begin{tikzpicture}
        \node[inner sep=0pt] 
        {\fbox{\includegraphics[width=\textwidth]{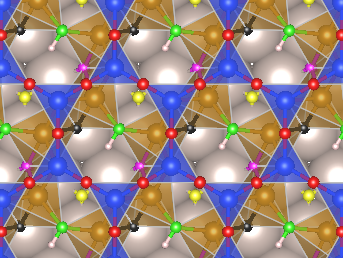}}};
        \node[align=center,fill=white,draw] at (-2.4, 1.8) {(b)};
        \node[coordinate] (A) at (0.85,0.7) {};
        \node[coordinate] (B) at (0.85,-0.2) {};
        \draw[double arrow=5pt colored by black and red,rounded corners] (A) -- (B);        
        \end{tikzpicture}
    \end{subfigure}
    \begin{subfigure}[htbp]{0.31\textwidth}
        \begin{tikzpicture}
        \node[inner sep=0pt] 
        {\fbox{\includegraphics[width=\textwidth]{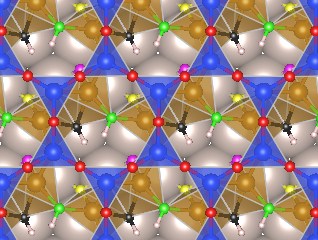}}};
        \node[align=center,fill=white,draw] at (-2.4, 1.8) {(c)};
        \node[coordinate] (A) at (0.55,0.65) {};
        \node[coordinate] (B) at (0.55,-0.25) {};
        \draw[double arrow=5pt colored by black and red,rounded corners] (A) -- (B);        
        \end{tikzpicture}
    \end{subfigure}
    \begin{subfigure}[htbp]{0.31\textwidth}
        \begin{tikzpicture}
        \node[inner sep=0pt] 
        {\fbox{\includegraphics[width=\textwidth]{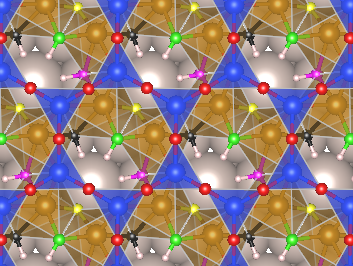}}};
        \node[align=center,fill=white,draw] at (-2.4, 1.8) {(d)};
        \node[coordinate] (A) at (0.97, 0.2) {};
        \node[coordinate] (B) at (0.97,-0.7) {};
        \draw[double arrow=5pt colored by black and red,rounded corners] (A) -- (B);        
        \end{tikzpicture}
    \end{subfigure}
    \begin{subfigure}[htbp]{0.31\textwidth}
        \begin{tikzpicture}
        \node[inner sep=0pt] 
        {\fbox{\includegraphics[width=\textwidth]{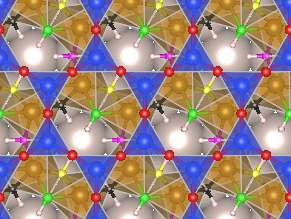}}};
        \node[align=center,fill=white,draw] at (-2.4, 1.8) {(e)};
        \node[coordinate] (A) at (1.4,1.6) {};
        \node[coordinate] (B) at (1.4,0.7) {};
        \draw[double arrow=5pt colored by black and red,rounded corners] (A) -- (B);        
        \end{tikzpicture}
    \end{subfigure} 
    \begin{subfigure}[htbp]{0.31\textwidth}
        \begin{tikzpicture}
        \node[inner sep=0pt] 
        {\fbox{\includegraphics[width=\textwidth]{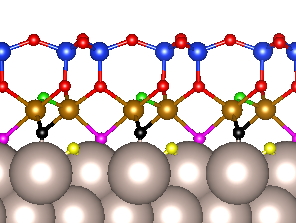}}};
        \node[align=center,fill=white,draw] at (-2.4, 1.8) {(f)};
        \node[text=green,rounded corners=2pt, inner sep=1pt] at (-0.1, 0.5) {\contour{black}{\bf  O$^{\rm c}$}};
        \node[text=yellow, rounded corners=2pt, inner sep=1pt] at (0.4, -0.45) {\contour{black}{\bf O$^{\rm ad}$}};
        \node[text=magenta, rounded corners=2pt, inner sep=1pt] at (-1, -0.25) {\contour{black}{\bf  O$^{\rm atop}$}};
        \node[text=black, rounded corners=2pt, inner sep=1pt] at (0.1, -0.15) {\contour{white}{\bf  O$^{\rm hcp}$}};
        \end{tikzpicture}
    \end{subfigure}        
\caption{Fe$_2$Si$_2$O$_9$H$_n\cdot$/Ru(0001) structures from $n=$ 0 to 4 in (a)-(e). In (a), the unit cell is shown with black lines. We indicate the oxygen atoms that can accommodate hydrogens in distinct sites in different colors. Corner sharing oxygen, O$^{\rm c}$, adsorbed oxygen O$^{\rm ad}$, oxygen on Ru atop site O$^{\rm atop}$ and oxygen on Ru hcp site O$^{\rm hcp}$ are indicated in green, yellow, magenta and black. Otherwise, the remaining oxygens make up Si-O tetrahedra and are indicated in red. Si, Fe, Ru, and H are indicated in blue, brown, gray, and white. In each figure (b)-(e), the additional hydrogens, from $n=1$ to $4$, are further indicated using a red arrow in the periodic cell. In (f) the side view of the structure in (a) is given.
}
\label{fgr:top_views}
\end{figure*}
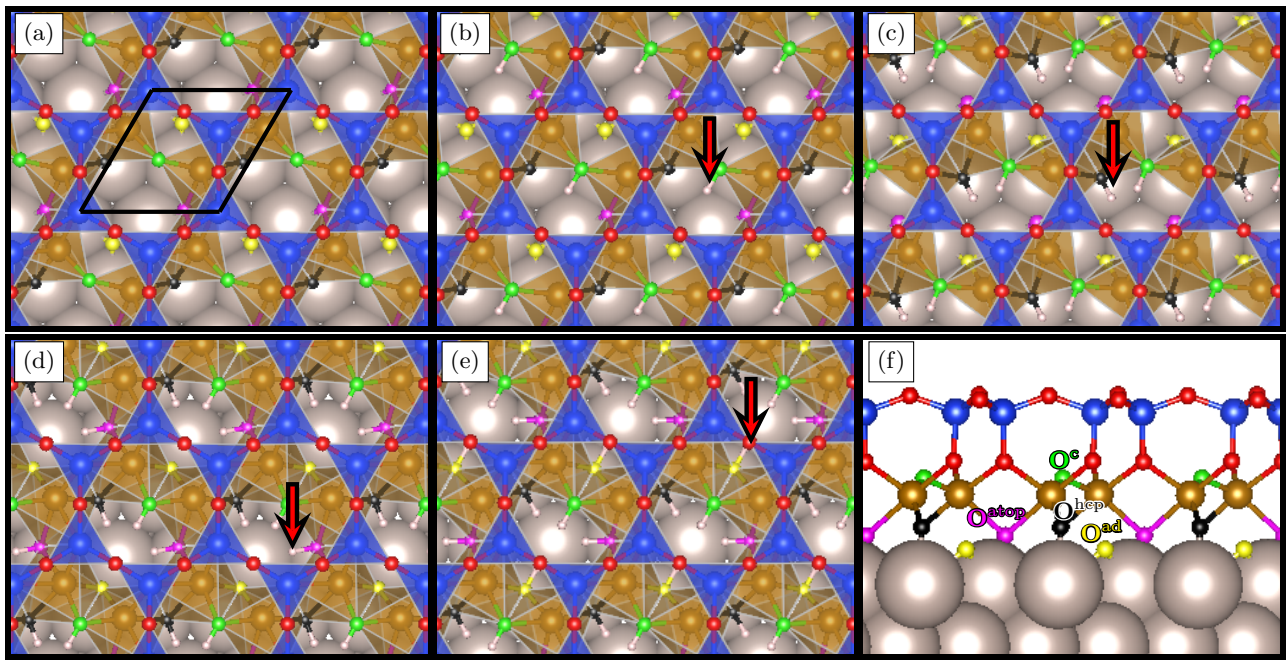

\section{Methods}
All DFT calculations were performed using the  plane-wave Vienna ab initio Simulation Package (VASP) \cite{Kresse1996, Kresse1996a}, using the Perdew-Burke-Ernzerhof exchange-correlation functional with D3 semi-empirical vdw correction (PBE+D3) \cite{Perdew1996, Grimme-d3}. We used a kinetic energy cutoff of 520 eV, periodic boundary conditions, projector-augmented wave (PAW) pseudopotentials \cite{Kresse1999} and dipole correction \cite{Neugebauer1992a}. We used Gaussian smearing of 0.2 eV to better capture the metallic nature of the film on the substrate together with a dense reciprocal grid density of 6$\times$6$\times$1. Geometry relaxations were performed until the forces on all atoms were smaller than 10$^{-2}$ eV/{\AA} with an electronic convergence of 10$^{-5}$ eV. Each simulation cell consists of a 2$\times$2 supercell of the 5-layer thick Ru (0001) surface primitive cell and a vacuum layer larger than 15 {\AA}. All the calculations were made with spin-polarization due to the presence of the Fe atom. All Fe atoms were initialized in the high-spin ferromagnetic configuration. 

Harmonic vibrational frequencies at $\Gamma$ were calculated using the frozen phonon method using central differences with 0.015 {\AA} displacements along with Cartesian directions. Zone-center phonon modes were obtained from the eigenvalues of the dynamical matrix. Experimentally, the RAIRS method is only sensitive to changes in the dipole moment vertical to the substrate surface. Therefore, simulated RAIRS intensities were obtained from the changes in the square of the dipole moments along the z-axis at each phonon mode \cite{Karhanek2010}. The vibrational frequencies were scaled with a factor of 1.0341 derived from a comparison between the theoretical and experimental vibrational frequencies of $\alpha$-quartz \cite{Loffler2010}. 

To calculate simulated surface core-level binding energy shifts (SCLS), ${\rm SCLS}$, we used the initial-state approximation via DFT\cite{Kohler2004}. SCLS can explain the relative positions of the core levels of an adsorbate, hence can be compared to XPS experiments.
In this approach, the core-level binding energy of an electron is first calculated as $E_B = - (\epsilon_{c} - E_F - \phi$) where $\epsilon_{c}$ is the Kohn-Sham eigenstate of the core level, $E_F$ and $\phi$ are Fermi level and the workfunction of the system. Therefore, ${\rm SCLS}$ are calculated as the difference in the core-level binding energies on substrate, ${E}_B^{\rm sub}$, and under vacuum, ${E}_B^{\rm vac}$: ${\rm SCLS} = {E}_B^{\rm sub} - {E}_B^{\rm vac}$ \cite{Lizzit}.
SCLS can occur due to multiple factors such as transitioning to a different oxidation state in the adsorbate due to charge transfer or modification of the interface dipole \cite{Egelhoff1987}. 
When comparing Fe-silicates with different amounts of hydrogens on Ru (0001), or under similar varying conditions, it is expected that the interface dipole is modified at every hydrogenation level.
We aim to understand how the charge states of Fe and O atoms change with increasing hydrogenation. Therefore, the effect of interface dipole on SCLS should be subtracted to the best approximation.
These considerations have also been observed in XPS studies where core levels of all species at the overlayer can shift depending on the strength of the interfacial dipole, which can introduce inconsistencies between measurements of different samples \cite{Jhang2020}.
For physisorbed silica bilayer on Ru(0001), XPS study shows that upon removal of the chemisorbed oxygen, O(2$\times$2)/Ru(0001), silica bilayer associated O-1$s$ and Si-2$p$ core levels are shifted uniformly by 0.75 eV, indicating that the effect of the interface dipole on the core level binding energies of adsorbates is uniform \cite{Wang2016b}.
Therefore, SCLS of a species with a well-defined oxidation state, such as Si$^{4+}$, can be taken as a reference across different samples each having different interface dipoles.
Si$^{4+}$ is a good reference, because Si$^{4+}$-p valence levels of Fe-silicates on Ru(0001) are energetically stable by nearly 20 eV below the Fermi level \cite{Supp} and we find that Si-$p$ valence states have a charge transfer of only 0.001 e from the Ru(0001) substrate. Therefore, the change in the Si-2$p$ core level shifts can be approximated to be primarily due to modification of the interfacial dipole, because the charge transfer is very limited.
Therefore, we use a modified definition of the SCLS for our oxidation state analysis ${\rm SCLS}' = {\rm SCLS} - \delta{\rm SCLS}^{\rm Si-2p}$ where $\delta{\rm SCLS}^{\rm Si-2p}= {\rm SCLS}^{\rm Si-2p} - {\rm SCLS}^{\rm Si-2p, ref}$ is the change in the SCLS of Si-2$p$ electrons compared to a reference system, which we choose as  Fe$_2$Si$_2$O$_8\cdot$O/Ru(0001). 
The correction due to $\delta{\rm SCLS}^{\rm Si-2p}$ anchors ${\rm SCLS}^{\rm Si-2p}$ at the reference ${\rm SCLS}^{\rm Si-2p, ref}$ energy across samples with varying amounts of hydrogens, hence the effect of interfacial dipole on the other core-levels is subtracted.
To simulate the ${\rm SCLS}'$ spectra we use Gaussians with a full half-width maximum of 0.8 eV  \cite{Jain2018, Baer2020}.

The chemical potential of the hydrogen gas at standard conditions is required to calculate the Gibbs formation enthalpies. Using DFT we calculate the H$_2$(g) dimer at 0 K and find that it has a dissociation energy of 4.54 eV compared to the experimental value of 4.48 eV \cite{darwent1970bond}. Then we write the reference chemical potential of H at standard conditions as, $\mu_{\rm H}^{\rm ref}{\rm({\rm T})} = 1/2 (E_{\rm H_2}^{\rm 0,DFT} - {\rm TS}_{\rm H_2}^{\rm exp}({\rm T}))$. Experimental entropy of H$_2$ gas under standard conditions, $S_{\rm H_2}^{\rm exp}({\rm T})$ is taken from NIST Webbook \cite{NIST} as 130.680 J/mol*K. Using this and the DFT energy of H$_2$ gas at 0 K, $E_{\rm H_2}^{0,DFT}$ of -6.774 eV, we find $\mu_{\rm H}^{\rm ref}({\rm T})$ under standard conditions as -3.571 eV. To define the hydrogen chemical potential under varying pressure at 298 K, we define $\mu_{\rm H}^{\rm ref}({\rm P, T=298K}) = \mu_{\rm H}^{\rm ref} - 1/2{\rm RTln}(\rm P_{H_2})$, where $\rm P_{H_2}$ is the hydrogen gas partial pressure under standard conditions. Surface binding energies are calculated from 
\begin{equation}
    E_b = E_A + E_{substrate} - E_{A/Substrate} 
\end{equation}
where $E_A$, $E_{substrate}$, and $E_{A/Substrate}$ are the total energies of isolated adsorbate, substrate, and the chemisorption system, respectively. In this definition, a positive binding energy $E_b$ means that the process is exothermic.

\section{Results and Discussion}
\label{sec:res}
We use Fe$_2$Si$_2$O$_8\cdot$O/Ru(0001) slab as the starting point of our theoretical investigation of its hydrogenation as shown in Fig. \ref{fgr:top_views}(a). This dehydroxylated system was previously studied by \citet{Wodarczyk2013} and here, we first reproduce their findings. As shown in Fig. \ref{fgr:top_views}, Fe$_2$Si$_2$O$_8\cdot$O/Ru(0001) is made of Fe-O layer at the interface with the substrate and has a nontronite-like as previously explained. Here, O/Ru(0001) stands for the pristine Ru(0001) surface plus an adatom O located at every other hcp site in (2$\times$2) pattern. This O/Ru(0001) phase is also known as the low coverage phase of adsorbed O on Ru (0001) \cite{Lizzit}. Therefore, Fe$_2$Si$_2$O$_8$ film is best understood as being bout to the O/Ru(0001) surface, hence combined making up the  Fe$_2$Si$_2$O$_8\cdot$O/Ru(0001) complex. Oxygens in the Fe$_2$Si$_2$O$_8\cdot$O/Ru(0001) complex can be separated into five groups depending on their positions and vertical distances to the Ru substrate: The adatom O, O$^{\rm ad}$, is closest to the Ru substrate and O$^{\rm ad}$ has a vertical distance, z(O$^{\rm ad}$-Ru), of 1.12 {\AA} to the top Ru layer. However, the the shortest Ru-O$^{\rm ad}$ bond, min(d(O$^{\rm ad}$-Ru)), is 2.03 {\AA}, . 
This 2.03 {\AA} d(O$^{\rm ad}$-Ru) bond length compares well to the experimental value of 1.99 {\AA} in RuO$_2$ \cite{Bolzan1997}.
Oxygens above the atop and hcp Ru sites, (O$^{\rm atop}$ and O$^{\rm hcp}$) as shown in Fig. \ref{fgr:top_views}, are further away from the substrate compared to O$^{\rm ad}$ and have an average z(O-Ru) of 1.8 {\AA}. 
The corner-sharing oxygens (O$^{\rm c}$) are, however, further distant to the Ru substrate by 3.57 {\AA}, and positioned at the shared corners of the FeO$_5$ polyhedra. The remaining oxygen atoms that are furthest away from the Ru substrate, by at least 4.2 {\AA}, make up the Si-O tetrahedra.
In the structure reported by \citet{Wodarczyk2013}, the O$^{\rm ad}$ is positioned below the Si-O-Si bridging oxygens along the $z$-axis. However, the Fe$_2$Si$_2$O$_8$ can slide over this substrate and can use a different registry. We construct our simulation using these constraints and slide the Fe$_2$Si$_2$O$_8$ along the substrate surface and find that the O adatom is positioned below the Si-O-Si is indeed the most stable configuration. We find that the binding energy of Fe$_2$Si$_2$O$_8$ on O/Ru(0001) substrate is 0.149 eV/{\AA}$^2$ (1.88 eV per Fe) which agrees nicely with the 0.160 eV/{\AA}$^2$ (-15.48  kJ$\cdot$mol$^{-1}$/{\AA}$^2$) based on the  DFT+D functional \cite{Grimme-d} \cite{Wodarczyk2013}.

\subsection{Binding of a single H on Fe$_2$Si$_2$O$_8\cdot$O/Ru(0001)}
\label{sec:single_h}
To study the effect of hydrogenation on the Fe$_2$Si$_2$O$_8\cdot$O/Ru(0001) system,  we first introduce a single hydrogen atom and investigate its binding energy to different oxygen sites. In the serpentine-silicate analogs such as nontronite, kaolinite, lizardite, and talc, the hydrogens are always found to be bonded to metal oxide groups rather than the Si-O tetrahedral groups.  Here, as well in our prior work \cite{Saritas2021}, we find this to be true as shown in Fig. \ref{fgr:top_views}(b). Therefore, we do not consider the oxygens that are part of Si-O network as potential hydrogen binding sites. Hence, out of a total 9 starting oxygen sites, we will study  hydrogen binding on the remaining 4 oxygen sites. These sites are, in the order of their distance to the Ru substrate, adsorbed oxygen O$^{\rm ad}$, oxygens above atop and hcp Ru (0001) sites, O$^{\rm atop}$ and O$^{\rm hcp}$, and finally the corner sharing Fe-O-Fe oxygen O$^{\rm c}$.  We study the binding energy of single H at a time separately to these four oxygen sites in this section. We find that the hydrogen binding energy in the Fe$_2$Si$_2$O$_8\cdot$O/Ru(0001) + 1/2 H$_2$ $\longrightarrow$ Fe$_2$Si$_2$O$_8\cdot$OH/Ru(0001) reaction is -1.22, -0.65, -0.25 and -0.07 eV for O$^{\rm c}$, O$^{\rm atop}$, O$^{\rm hcp}$ and O$^{\rm ad}$ respectively. It is interesting to note that the H-binding energy at O$^{\rm c}$ site is significantly larger than the binding at other sites considered. The distinctly lower hydrogen binding energy at the O$^{\rm c}$ site was also observed at the nontronites under vacuum \cite{Saritas2021}. This will be further discussed in Sec. \ref{sec:comp_hyd}.

The difference in the hydrogen binding energies at O$^{\rm atop}$ and O$^{\rm hcp}$ sites can be understood via the mechanism of water dissociation at Ru (0001) surfaces \cite{Michaelides2003}. Water on  Ru (0001) is known to partially dissociate into H and OH at 100-170 K \cite{Feibelman2002}. Its structure has been characterized experimentally \cite{Feibelman2002}, and theoretical research confirmed this structure with a dissociation barrier of 0.5 eV which agrees with the experimental dissociation temperature \cite{Michaelides2003, Tatarkhanov}. Water dissociation on Ru (0001) is energetically quite favorable with an enthalpy of -0.85 eV per dissociated H$_2$O \cite{Michaelides2003}. Theoretical calculations find those intact H$_2$O molecules, before their eventual dissociation on Ru (0001), form a bilayer with two types of molecules: one nearly parallel to the surface and the second with an O-H bond nearly perpendicular to the surface. However, it is thermodynamically favorable to dissociate every one of two water molecules on the substrate. Upon the dissociation, the OH, H, and H$_2$O all bind parallel to the surface, hence forming a monolayer, and are adsorbed on atop Ru (0001) sites \cite{Michaelides2003}. OH and H$_2$O form a honeycomb lattice by forming  hydrogen bonds parallel to the substrate, whereas the H atoms are positioned in the center of each honeycomb site \cite{Feibelman2002}. Based on this, in the case of Fe$_2$Si$_2$O$_8\cdot$OH/Ru(0001), H binding on O$^{\rm atop}$ should be more favorable because of the collective preference of H$_2$O and OH groups to be located at the atop Ru sites. For OH and H$_2$O on Ru (0001), it was noted that the surface adsorption mainly occurs thanks to an H-$s$/O-$p$ hybridized orbital and Ru-$d_{z^2}$ orbitals when H$_2$O is located at favorable atop-site \cite{Michaelides2003}. When the OH or H$_2$O is not located on an atop Ru site, this overlap may not be as strong as in the case of atop case, hence leading to the difference in H-binding energies at O$^{\rm atop}$ and O$^{\rm hcp}$ sites. 

\begin{figure}[t]
\centering
    \includegraphics[width=\linewidth]{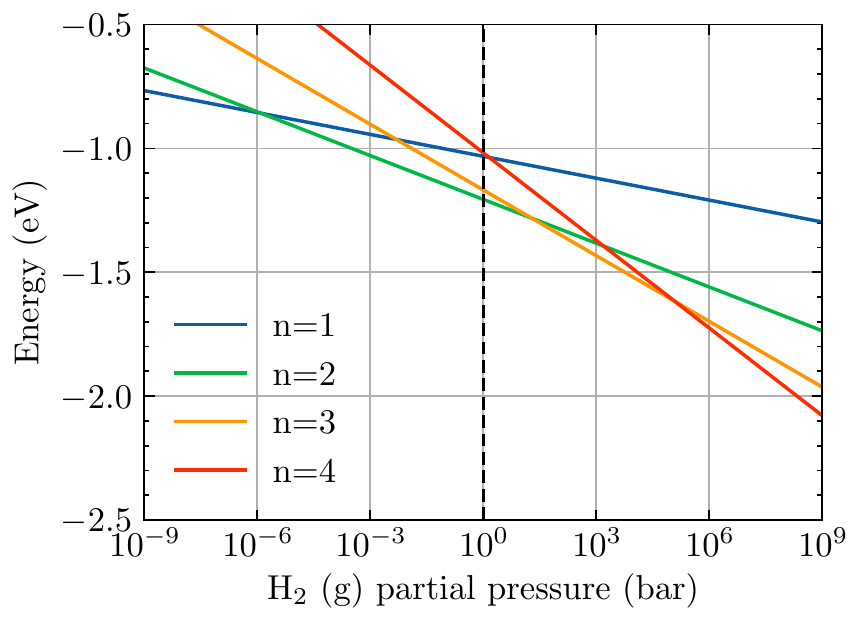}
  \caption{Gibbs free energy of formation at 298 K, $\Delta${G}, for the Fe$_2$Si$_2$O$_9$/Ru(0001) + n/2 H$_2$ $\longrightarrow$ Fe$_2$Si$_2$O$_9$H$_n$/Ru(0001) reaction}
  \label{fgr:gibbs}
\end{figure}

The reason for low hydrogen binding energy at O$^{\rm ad}$ can also be explained by similar reasoning of the difference in hydrogen binding energies at O$^{\rm atop}$ and O$^{\rm hcp}$. After binding hydrogen to O$^{\rm ad}$, we find that the geometry relaxation yields a structure where the O$^{\rm ad}$ migrates from its original hcp position towards a bridge site between two top layers of Ru atoms.
We tried various orientations of the O$^{\rm ad}$-H bond to the substrate surface and found that the vertical orientation of O$^{\rm ad}$-H group towards the substrate is not energetically favorable. Relaxed geometry of O$^{\rm ad}$-H group is angled such that its H atom forms a hydrogen bond with the O$^{\rm atop}$ with a d(O$^{\rm atop}$-H) distance of 1.96 {\AA}. This is similar to the partially dissociated water on Ru (0001), however, the migration of the O$^{\rm ad}$ to an atop site can be hindered by the presence of the interface oxygen atoms of the Fe-silicate layer. This indicates that  Fe$_2$Si$_2$O$_8\cdot$O/Ru(0001) has a similar complexity for hydrogen bonding networks which facilitates a particular orientation of the OH groups. 

\begin{figure}[t]
\centering
    \includegraphics[width=\linewidth]{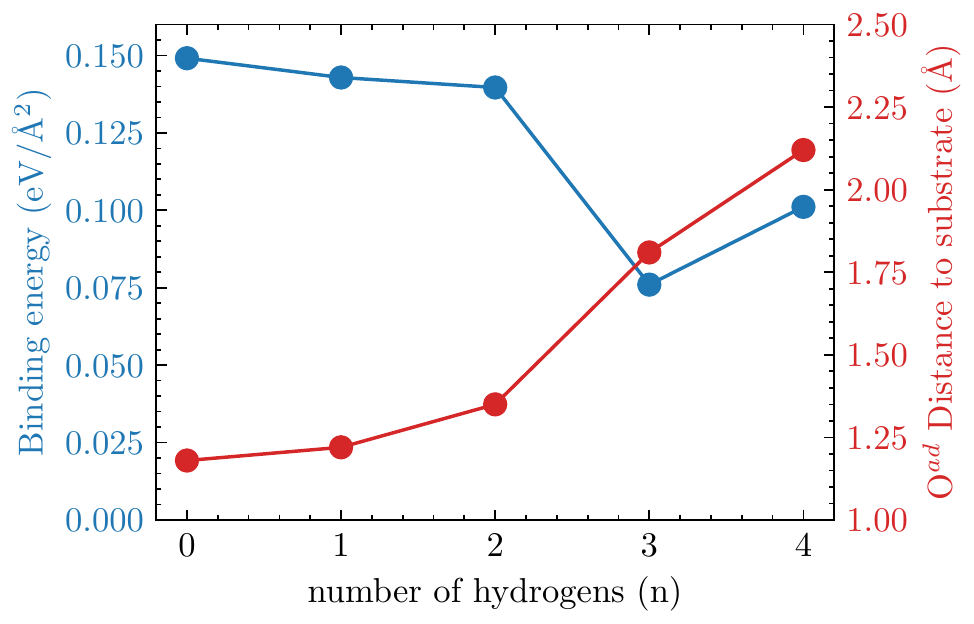}
  \caption{The binding energies of each of these compounds and the distance of the adatom oxygen O$^{\rm ad}$ to the substrate surface are given.}
  \label{fgr:binding}
\end{figure}

\subsection{Complete hydrogenation of Fe$_2$Si$_2$O$_8\cdot$O/Ru(0001) to Fe$_2$Si$_2$O$_9$H$_4$/Ru(0001)}
\label{sec:comp_hyd}
Having studied how a single H atom  bonds to  Fe$_2$Si$_2$O$_8\cdot$O/Ru(0001), we now consider further hydrogenation. We considered up to four H bound to the four favorable sites of Fe$_2$Si$_2$O$_8\cdot$O/Ru(0001) as explained in the previous section. At each hydrogenation level, we considered all possible combinations of different sites. We find that the qualitative energetic ordering of the binding energies for single hydrogen adsorption applies while achieving complete hydrogenation. In Fig.~\ref{fgr:gibbs}, we plot the Gibbs free energies of hydrogenation for the
\[
{\rm Fe}_2{\rm Si}_2{\rm O}_9/{\rm Ru}(0001) + \frac{n}{2} H_2 \longrightarrow {\rm Fe}_2{\rm Si}_2{\rm O}_9{\rm H}_n/{\rm Ru}(0001)
\]
reaction and show that the formation of Fe$_2$Si$_2$O$_9$H$_2$ is thermodynamically favorable under ambient conditions. According to Fig. \ref{fgr:gibbs}, higher H$_2 (g)$ pressures in the hundreds of kbar range are needed to completely hydrogenate Fe$_2$Si$_2$O$_9$H$_4$. Alternatively, one can consider H atoms generated using a plasma source or an H atom source \cite{Tschersich2008}.

Introducing more than single hydrogen to Fe$_2$Si$_2$O$_8\cdot$O/Ru(0001) causes some changes in the geometry of the complex which facilitates gradual migration of O$^{\rm ad}$ towards the Fe-silicate film from the Ru substrate. 
With the addition of second hydrogen ($n=2$) at O$^{\rm atop}$ site (the second most favorable hydrogen binding oxygen site),  shown in Fig. \ref{fgr:top_views}(c), one of the Fe atoms move towards the O$^{\rm ad}$. Initially at the dehydroxylated case, $n=0$, the closest d(Fe-O$^{\rm ad}$) distances are 2.67 and 2.81 {\AA}. However, at $n=2$ one d(Fe-O$^{\rm ad}$) is 2.07 {\AA} while the other becomes 2.91 {\AA}. 
Average Ru-O$^{\rm ad}$ bond lengths also increase from 2.01 to 2.10 {\AA} as $n$ increases from 0 to 2.
Therefore, these collective changes can indicate that Ru-O$^{\rm ad}$ bonds are weakened or Fe-O$^{\rm ad}$ bonds got stronger with hydrogenation.
To understand whether O$^{\rm ad}$ is more strongly bound to the substrate or the Fe-silicate at $n=2$, we explore two different scenarios: We find that the Fe$_2$Si$_2$O$_8$H$_2\cdot$O/Ru(0001) $\longrightarrow$ Fe$_2$Si$_2$O$_8$H$_2$ + O/Ru(0001) reaction yields a surface binding energy of 1.76 eV/Fe (0.14 eV/{\AA$^2$}), whereas the Fe$_2$Si$_2$O$_9$H$_2\cdot$O/Ru(0001) $\longrightarrow$ Fe$_2$Si$_2$O$_9$H$_2$ + Ru(0001) reaction has a surface binding energy of 2.62 eV/Fe (0.21 eV/{\AA$^2$}). Therefore, at $n=2$, the Ru-O$^{\rm ad}$ bonds are harder to break than the Fe-O$^{\rm ad}$ bonds, hence O$^{\rm ad}$ still prefers being on the substrate rather than leaving the substrate with the film upon exfoliation.
Therefore, we conclude that the addition of only two hydrogens per formula unit Fe-silicate to this slab still yields very large binding energy between the overlayer and substrate. 

\begin{figure}[t]
\centering
    \begin{tikzpicture}
        \node[inner sep=0pt] 
        {\includegraphics[width=\linewidth]{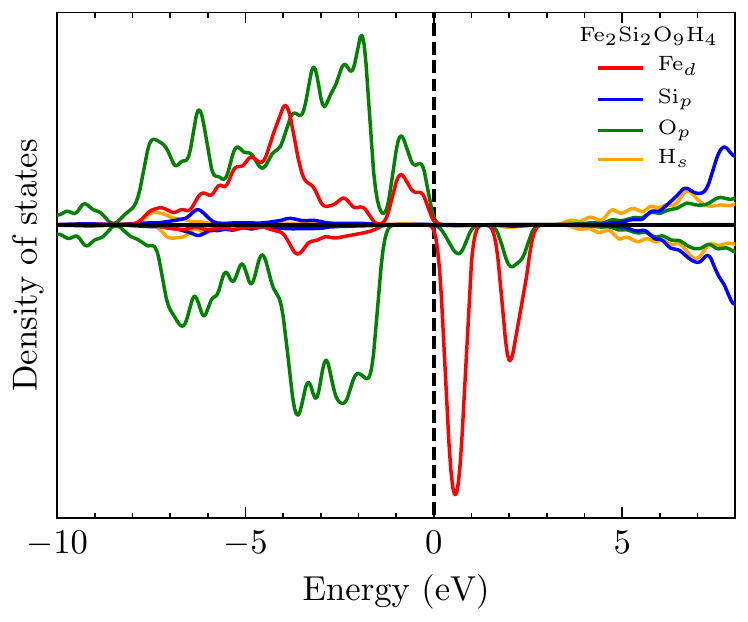}
};    
    \node[align=center,fill=white,draw] at (-3.2, 2.9) {(a)}; 
    \end{tikzpicture}
    \begin{tikzpicture}
        \node[inner sep=0pt] 
        {\includegraphics[width=\linewidth]{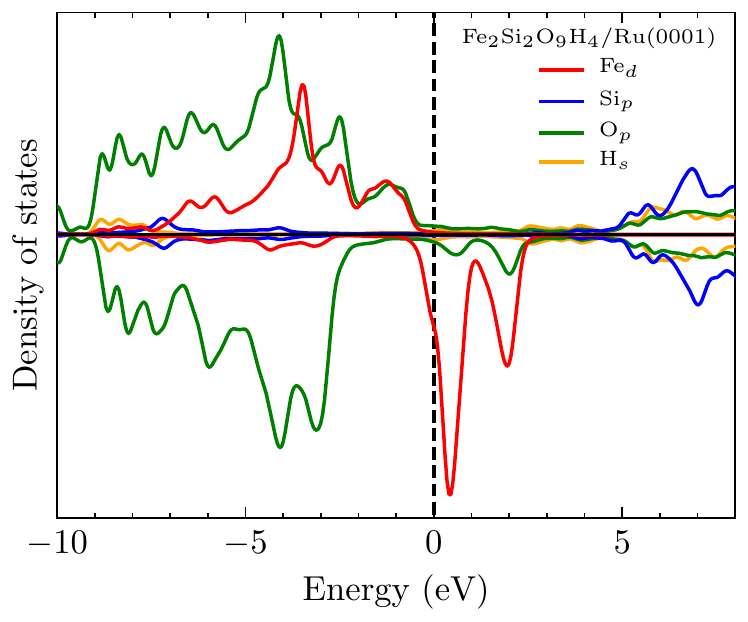}};    
    \node[align=center,fill=white,draw] at (-3.2, 2.9) {(b)}; 
    \end{tikzpicture}
    
  \caption{Atom decomposed densities of states of (a) Fe$_2$Si$_2$O$_9$H$_4$ in vacuum and (b) Fe$_2$Si$_2$O$_9$H$_4$/Ru(0001). }
  \label{fgr:dos}
\end{figure}

The third hydrogen can be bonded to the O$^{\rm hcp}$, as shown in Fig. \ref{fgr:top_views}(d). Therefore, the O$^{\rm hcp}$-H group migrates towards an atop Ru site by 0.8 \AA, while the Si-tetrahedra slide about 1.1 {\AA} on the film in the same direction as the O$^{\rm hcp}$-H group. This causes a significant reorganization for O$^{\rm ad}$ atom which moves from its original hcp site to a bridge site between two top Ru atoms while (O$^{\rm ad}$-Ru) increases from 1.35 {\AA} to 1.80 \AA. To explain these significant changes at the $n=3$ step, we plot the surface binding energies and z(O$^{\rm ad}$-Ru) as a function of hydrogenation level in Fig. \ref{fgr:binding}. As shown in Fig. \ref{fgr:binding}, the vertical distance of the  O$^{\rm ad}$ to the Ru substrate, z(O$^{\rm ad}$-Ru), increases as 1.18, 1.22, 1.35, 1.80, 1.98 \AA with increasing the number of hydrogens, $n$, from 0 to 4. Therefore, there is a trend of O$^{\rm ad}$ moving away from the substrate even before this significant reorganization at $n=3$ happens. Up to $n=2$, we see no change in the lateral position of the O$^{\rm ad}$, meaning that its Ru(0001)-hcp site is still strongly favorable for O$^{\rm ad}$ and O$^{\rm ad}$ is most strongly interacting with the Ru substrate. Similar to the $n=2$ case, to quantify whether the O$^{\rm ad}$ would like to bond to the film or the substrate at $n=3$, we again compare two different scenarios: We find that the Fe$_2$Si$_2$O$_9$H$_3$ on Ru(0001) has a surface binding energy of 0.149 eV/{\AA$^2$}, whereas Fe$_2$Si$_2$O$_8$H$_3$ on O/Ru(0001) has a surface binding energy of 0.076 eV/{\AA$^2$}. Therefore, at $n=3$ hydrogenation level, the O$^{\rm ad}$ adsorption on Ru significantly weakens. However, according to Fig. \ref{fgr:gibbs}, hydrogen  gas partial pressures of near 1 kbar or with hydrogen richer conditions/sources are needed to achieve Fe$_2$Si$_2$O$_9$H$_3$/Ru(0001). 

With the addition of the fourth hydrogen to the O$^{\rm ad}$ site, as shown in Fig. \ref{fgr:top_views}(e), the vertical distance of this O$^{\rm ad}$ atom to the Ru substrate increases to 1.98 \AA. This is a substantial increase over the 1.18 {\AA} z(O$^{\rm ad}$-Ru) at $n=0$.  Fe$_2$Si$_2$O$_9$H$_4$ film on the Ru substrate has a binding energy of 1.27 eV/Si (0.10 eV/{\AA$^2$}). Interestingly, this indicates an increase over the binding energy at $n=3$ case. At this point, it should be noted that hydrogen gas partial pressures of 10$^5$ bar are needed to thermodynamically stabilize Fe$_2$Si$_2$O$_9$H$_4$/Ru(0001), which indicates about four orders of magnitude richer H gas richer conditions compared to the onset of Fe$_2$Si$_2$O$_9$H$_3$/Ru(0001) formation according to Fig. \ref{fgr:gibbs}. The 0.10 eV/{\AA$^2$} binding energy suggests that the interaction is not fully a van-der Waals type but includes a degree of chemical adsorption. Nevertheless, materials with up to 0.15 eV/{\AA$^2$} binding are often regarded as potentially exfoliable, therefore it may be possible to extract a 2D Fe-silicate layer  using special techniques \cite{Gibertini2019}. 

\begin{figure}[t]
\centering
    \begin{tikzpicture}
        \node[inner sep=0pt] 
        {\includegraphics[width=\linewidth]{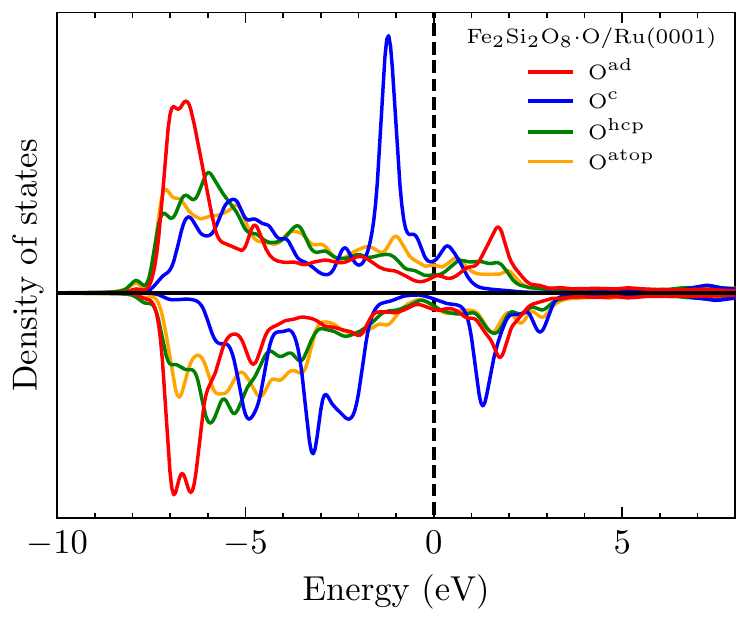}
};    
    \end{tikzpicture}
  \caption{Partial densities of states of the $p$ orbitals of O$^{\rm ad}$, O$^{\rm c}$, O$^{\rm hcp}$ and O$^{\rm atop}$ at Fe$_2$Si$_2$O$_8\cdot$O/Ru(0001) structure. }
  \label{fgr:oxydos}
\end{figure}

To explain the differences in the hydrogen binding energies at different oxygen sites, it can be useful to study the changes in densities of states of Fe-silicates upon hydrogenation. Therefore, in Fig. \ref{fgr:oxydos}, we show the detailed partial densities of states of hydrogen binding oxygens at Fe$_2$Si$_2$O$_8\cdot$O/Ru(0001) structure. The sharp atomic-like O$^{\rm c}-p$ peak is close to Fermi energy at about -1 eV and the large O$^{\rm ad}-p$ peak is at -7 eV are the contrasting features of these two oxygens. 
We find that upon complete hydrogenation at $n=4$ \cite{Supp}, the sharp atomic-like O$^{\rm c}-p$ peak reduces in energy by about 6-7 eV. This is in contrast to other O-$p$ peaks which show much smaller or zero reductions. In Fig. \ref{fgr:oxydos}, O$^{\rm ad}-p$ states are centered at the lowest energy level compared to other hydrogen-binding oxygens, which would mean that the O$^{\rm ad}$ is passivated more strongly.  Upon complete hydrogenation at $n=4$, however, partial densities of states of O$^{\rm atop}$ and O$^{\rm hcp}$ become exactly equal. 

\begin{table}[t]
\begin{tabular}{c|ccccc}
\hline
\hline
  property   & $n=$0 & $n=$1 & $n=$2 & $n=$3&$n=$4 \\
\hline     
${\Delta}$E (meV/{\AA}$^2$) & 0.149 & 0.142 & 0.139 & 0.076 & 0.101\\
${\Delta}\Phi$ (eV) & 1.02 & 1.02 & 0.46 & 0.95 & 0.72 \\
z(O$^{\rm ad}$-Ru) ({\AA}) & 1.18 & 1.22 & 1.35 & 1.81 & 1.98\\
${\Delta\rho}$ (e) & 0.75 & 0.71 & 0.46 & 0.95 & 0.76 \\
\hline
\end{tabular}
\caption{The binding energy of the adlayer(${\Delta}$E), work function change with respect to the O/Ru(0001) substrate (${\Delta}\Phi$), vertical separation of the O$^{\rm ad}$ from the Ru layer (z(O$^{\rm ad}$-Ru)), Bader charge of the adsorbed layer ${\Delta\rho}$).}
\label{tab:tab}
\end{table}

\subsection{Changes in the work function with increasing hydrogenation}
Hydrogenation of the Fe-silicate can affect the formation of the interface dipole, and the amount of charge transfer between the substrate and the film. Therefore, the workfunction of the substrate-film complex can be modified. First, we calculate that the bare Ru (0001) surface has a work function of $\Phi$=5 eV compared to an experimental value of 5.4 V \cite{Bo, Himpsel1982}. The Fe$_2$Si$_2$O$_8\cdot$O/Ru(0001) complex has a work function of 6.02 eV, which indicates a change in the workfunction, $\Delta\Phi$, by about 1.02 eV which is close to 1.4 eV from \citet{Wodarczyk2013}.  Our results of the changes in the work functions from bulk, $\Delta\Phi$, with increased hydrogenation of the Fe-silicate are given in Table \ref{tab:tab}.
$\Delta\Phi$ initially decreases as a function of hydrogenation from 1.02, 1.02 to 0.46 eV for $n$=0 to 2, indicating that the charge transfer is significantly decreasing as a function of hydrogenation. However, at $n$=3 the film reorganizes and incorporates the O$^{\rm ad}$, hence the $\Delta\Phi$ again increases to 0.95 eV. However, for complete hydrogenation at $n=4$, the $\Delta\Phi$ again reduces to 0.72 eV meaning the charge transfer in Fe$_2$Si$_2$O$_8$H$_2$ and Fe$_2$Si$_2$O$_9$H$_4$ are likely similar. As shown in Table \ref{tab:tab}, changes in $\Delta\Phi$ correlate with the amount of charge transfer. This is expected considering Fe has a formal charge of 3+ in both compounds if they were under vacuum, but a positive change in the workfunction would still indicate that there is charge transfer from the substrate to the film. In the SI \cite{Supp}, we show that the work function of bare Fe$_2$Si$_2$O$_9$H$_4$ in a vacuum is about 1.5 eV larger than the work function of Ru substrate, hence a charge transfer from the Ru substrate to the overlayer would be expected even when the overlayer is completely hydrogenated.

\subsection{Ionic states of Fe and O with increasing hydrogenation}
To further guide experimental synthesis and characterization, we compute additional observables.  First, we try to determine the charge state of the Fe ion as a function of the hydrogenation level. Therefore, using LOBSTER code \cite{Maintz2016, Dronskowski1993} we project the plane-wave wavefunction onto atomic orbitals and study integrated densities of states. We find that the improved basis sets LOBSTER can yield numerically more accurate results for quantities such as integrated densities of states, compared to the internal routines in VASP \cite{Maintz2013}.

To understand how the oxidation states of Fe-silicates evolve on Ru, we first explore Fe$_2$Si$_2$O$_9$H$_4$ in a vacuum. In a vacuum, Fe$_2$Si$_2$O$_9$H$_4$ would have no way to exchange charge with its surroundings, hence the simple assignment of formal charges to $sp-$block elements yields a 3+ charge for Fe. 
In Fig. \ref{fgr:dos}a, we plot the partial densities of states (DOS) for Fe$_2$Si$_2$O$_9$H$_4$ in a vacuum and show that simply enforcing the octet rule for $sp$-block elements yields the correct oxidation state for Fe atoms. 
We enforce high-spin starting conditions for Fe in all the calculations. As shown in Fig. \ref{fgr:dos}a, the high-spin Fe orbitals are completely occupied. 
We integrate Fe-$d$ curves up to Fermi level and find 4.98/0.86 (spin-up/down) $d$ electrons per Fe for Fe$_2$Si$_2$O$_9$H$_4$ in vacuum.
However, using the formal charge argument, meaning Fe would be 3+ in Fe$_2$Si$_2$O$_9$H$_4$, would indicate that Fe$^{3+}$ is $d^5$, hence would have no spin-down electrons.
The fact that the down spin O-$p$ electrons are also polarized above the Fermi level and the Fe-$d$ up spin orbitals are full indicates that the O-$p$ and Fe-$d$ are mainly hybridized in the minority spin channel and Fe-$d$ down spin $d$ electrons in Fig. \ref{fgr:dos}a formally belong to oxygen atoms. 
We support our findings using crystal orbital Hamilton population analysis (COHP)\cite{Dronskowski1993} in the SI \cite{Supp} which shows that Fe-O bonding is over the minority spins.

We analyze the partial DOS of Fe$_2$Si$_2$O$_9$H$_4$ on Ru and find that the Fe and O densities are very similar to the vacuum case, hence the integrals up to the Fermi level are 4.96/0.97 (spin-up/down).
In a vacuum, down spin Fe-$d$ orbitals are centered around -4 eV, while on Ru, this density is concentrated around the Fermi level, which could indicate that Fe atoms interact with the substrate Ru atoms, through O, as the work function of Ru is smaller than Fe$_2$Si$_2$O$_9$H$_4$. The DOS in Fig. \ref{fgr:dos}a, indicates that Fe$_2$Si$_2$O$_9$H$_4$ is a semi-metal. Therefore, a discussion on the choice of the DFT functional can be appropriate. With additional exchange splitting through Hubbard-$U$, an electronic gap would open in Fig. \ref{fgr:dos}a, which would make Fe$_2$Si$_2$O$_9$H$_4$ in a vacuum insulating with a finite gap. Nevertheless, PBE still yields a sizeable exchange splitting of nearly 5 eV and as the spin-up energy levels of Fe-$d$ electrons are completely occupied for under vacuum and on Ru (0001), results from PBE and PBE+$U$ would be qualitatively consistent. PBE functional was used to simulate Ti-silicate \cite{Fischer2015} and Fe-silicate \cite{Wodarczyk2013} on Ru (0001) previously and yield accurate structural and vibrational properties.

For the lesser hydrogenated Fe-silicates, however, the situation can be more complicated. For the completely dehydroxylated case, Fe$_2$Si$_2$O$_8\cdot$O/Ru(0001), assuming the substrate is defined as O/Ru(0001), the overlayer would be Fe$_2$Si$_2$O$_8$. For Fe$_2$Si$_2$O$_8$ in a vacuum, the formal charge assignment method would yield an oxidation state of 4+ on Fe. However, we find that as $n$ goes from 0 to 4 for Fe$_2$Si$_2$O$_{8,9}$H$_n\cdot$O$_{1,0}$/Ru(0001), the partial densities of states on Fe changes very weakly such that the total number of electrons on each Fe atom remain between 5.93 and 6.04 $e$ calculated using their partial DOS \cite{Supp} 
This highlights the difficulties in assigning an integer charge state to transition metal atoms due to charge self-regulation \cite{Raebiger2008}.

\subsection{Surface core-level binding energy shifts}
\label{sec:core}
While it would be useful to assign integer oxidation states to the Fe atoms at different hydrogenation levels of $n$, it is more important to relate the structural and physical changes to the observables from the experiments. 
For this reason, we calculate surface core-level binding energy shifts (SCLS) using DFT which can directly be compared to experimental x-ray photoelectron spectroscopy (XPS) results. 
SCLS can arise from two main effects: electrostatic potential, hence the chemical environment of a species, and its oxidation state \cite{Fongkaew2017}. Both of these factors can modify the screening of the core electrons, hence leading to varying binding energies. Therefore, chemical changes in a compound, such as an adsorbate on a substrate, can be understood through the SCLS of its species.  In this section, we investigate the SCLS' of Fe and O at various hydrogenation levels of the iron silicate films on Ru(0001) to understand the changes in the oxidation state of Fe ions. 

\begin{figure}[t]
\centering
    \includegraphics[width=\linewidth]{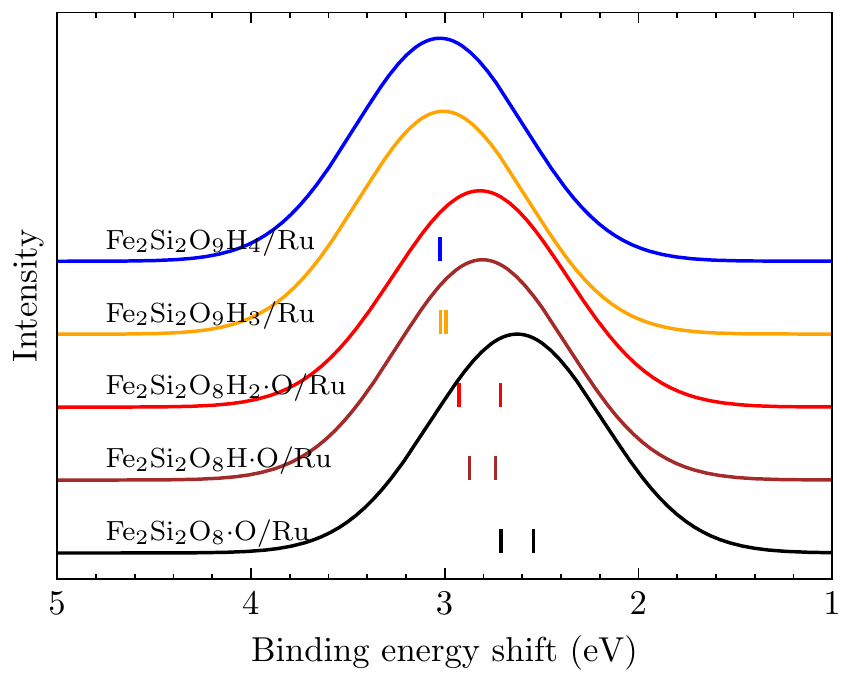}
  \caption{Fe-2$p$ surface core level binding energy shifts, SCLS', with increasing hydrogenation in Fe-silicate films on Ru(0001).  Short vertical lines indicate the discrete core level energies from DFT, which are used to produce the curves.}
  \label{fgr:xps-Fe}
\end{figure}

Fig. \ref{fgr:xps-Fe} shows that the SCLS of Fe atoms increases with the hydrogenation level, $n$. Additionally, it can be seen in Fig. \ref{fgr:xps-Fe} that up to $n=2$, Fe SCLS' are split by 0.25 eV and for $n=4$ the energies become completely degenerate. The energy split could be due to the difference in their registries to substrate. We find that the distances of Fe atoms to the O$^{\rm ad}$ are 2.81 and 2.64 {\AA}. These differences in surface registries can cause different screening properties. Once the O$^{\rm ad}$ is lifted off from the substrate with increased hydrogenation at $n=3$ and minimum z(O-Ru) distance is about 1.81 {\AA}, this splitting  also largely disappears as the effect of surface registry diminishes for Fe atoms. SCLS' of Fe can explain how the oxidation state of Fe changes with increased hydrogenation. Given that the oxidation states of Fe in Fe$_2$Si$_2$O$_9$H$_4$ under vacuum and Fe$_2$Si$_2$O$_9$H$_4$/Ru should be very close and can be considered close to 3+, the decreased SCLS' observed in Fig. \ref{fgr:xps-Fe} means that the Fe in Fe$_2$Si$_2$O$_8\cdot$O/Ru should be oxidizing towards 4+ states \cite{Pasquarello1995}. The difference in the Fe SCLS' in $n=0$ and $n=4$ cases is nearly about 0.4 eV. We can compare our results to bulk La$_{1-x}$Sr$_x$FeO$_3$, where a transition between 4+/3+ can be observed depending on the value of $x$ \cite{Wang2019}. For $x=0$ case in La$_{1-x}$Sr$_x$FeO$_3$, Fe is formally 3+. At $x=1$ Fe-2p binding energies shift to higher energies by about 1 eV \cite{Wang2019}. The 1 eV shift here is larger compared to our finding of nearly 0.4 eV shift between $n=0$ and $n=4$ cases of Fe silicate on Ru. However, an exact quantitative agreement would require more detailed calculations which involve final state and dynamic screening effects, hence our calculations are useful to provide qualitative results.

\begin{figure}[t]
\centering
    \includegraphics[width=\linewidth]{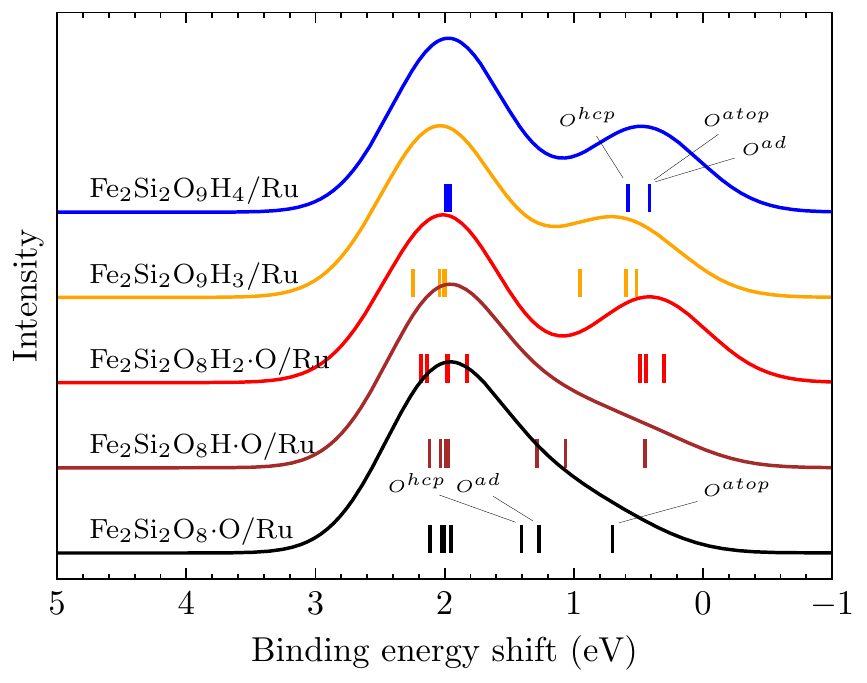}
  \caption{O-2$p$ surface core level binding energy shifts, SCLS', with increasing hydrogenation in Fe-silicate films on Ru(0001).  Short vertical lines indicate the discrete core level energies from DFT, which are used to produce the curves.}
  \label{fgr:xps-O}
\end{figure}

In Fig. \ref{fgr:xps-O} we show the core level shifts in oxygen atoms with increased hydrogenation. 
Our results in Fig. \ref{fgr:xps-O} for Fe$_2$Si$_2$O$_8\cdot$O/Ru(0001) can be compared to previous experiments \cite{Wodarczyk2013,Loffler2010} and theory \cite{Yang2012} on Fe silicates on Ru and bilayer silicates on Ru.
Our results here agree with these previous findings such that there are distinct SCLS' for O$^{\rm hcp}$, O$^{\rm atop}$ and O$^{\rm ad}$ which result in a shoulder formation in the O-1$s$ XPS with a nearly 2 eV shift compared to the large O-1$s$ peak that is associated with Si-O oxygens. Overall, the SCLS' around 2 eV in Fig. \ref{fgr:xps-O} are rather stationary with increased hydrogenation, meaning that oxygens associated with these peaks have constant charge transfer with increased hydrogenation. However, the SCLS' of chemisorbed oxygens (O$^{\rm hcp}$, O$^{\rm atop}$ and O$^{\rm ad}$) typically decrease with increased hydrogenation. For these oxygens, the trend is opposite that of Fe-2$p$ SCLS' in Fig. \ref{fgr:xps-Fe}. These O atoms have decreasing charge transfer from the Ru substrate with increased hydrogenation.
This is also correlated to the increasing z(O-Ru) with increased hydrogenation. 
While the O$^{\rm ad}$ are O$^{\rm atop}$ 1$s$ SCLS' are distinct in Fe$_2$Si$_2$O$_8\cdot$O/Ru(0001), they become isoenergetic for Fe$_2$Si$_2$O$_9$H$_4$/Ru(0001). 
This can also be related to the fact that with increased hydrogenation, the distance between the film and the substrate increases, and the effect of the surface registry decreases. However, the fact that O$^{\rm hcp}$ core level shift is different than O$^{\rm atop}$ and O$^{\rm ad}$ shows that there is still some effect of the surface registry. While the SCLS' of Fe$_2$Si$_2$O$_8\cdot$O/Ru(0001) can be seen as a peak and a small shoulder forming at lower binding energies, at the limit of complete hydrogenation, Fe$_2$Si$_2$O$_9$H$_4$/Ru(0001), it can be seen as a combination of two peaks with more distinct energies. Additional hydrogens increase the z(Ru-O) distances and decrease the ionic character of interface oxygens, which can lead to reduced SCLS'. SCLS' of oxygens around Si-O tetrahedra and O$^{\rm c}$ remain largely fixed as these oxygens are already well separated from the substrate and interact with the substrate only indirectly. 

\subsection{Simulated RAIRS}
\label{sec:rairs}
\begin{figure}[t]
{\centering
\includegraphics[width=\linewidth]{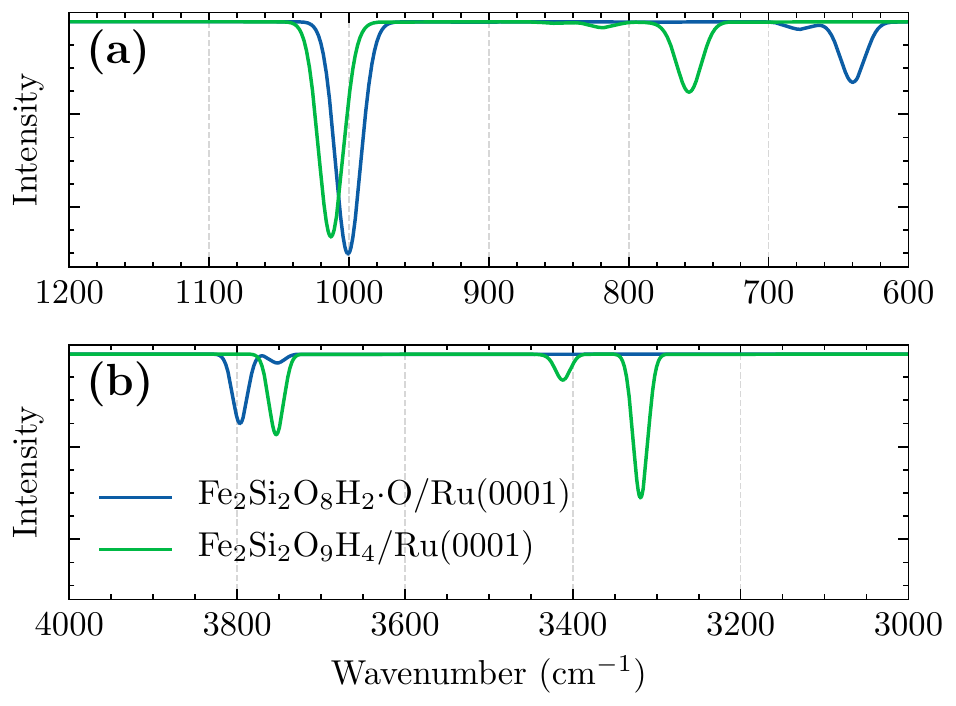}
 } \caption{Simulated RAIRS for the Fe$_2$Si$_2$O$_9$H$_4\cdot$/Ru(0001) and Fe$_2$Si$_2$O$_8$H$_2\cdot$O/Ru(0001) structures. The spectra is scaled by a factor of 1.0341 along the $x$-axis. Spectrum (a) shows the wave number range 600-1200 cm$^{-1}$, and (b) 3000-4000 cm$^{-1}$; between 1200 and 3000 cm$^{-1}$, no features are found. For (b), the intensity ($y$ axis) has been enlarged by 100x compared to (a) for better visibility.}
  \label{fgr:rairs}
\end{figure}

As increased hydrogenation is accompanied by structural changes in the Fe-silicate film, a change of the vibrational spectrum should be observed as the Fe-O layer goes from a 5-fold edge/corner-sharing network to an octahedral edge-sharing network with increased hydrogenation.  RAIRS experiments have shown that the Fe$_2$Si$_2$O$_8\cdot$O/Ru(0001) has a large peak at 1005 cm$^{-1}$, and through DFT calculations, this peak is associated with the vertical vibrations of bridging Fe-O-Si oxygens \cite{Wodarczyk2013}. In Fig. \ref{fgr:rairs}(a), we show that our Fe$_2$Si$_2$O$_8$H$_2\cdot$O/Ru(0001) also has a peak at 1002 cm$^{-1}$ which is due to the vertical motion of the Fe-O-Si bridging oxygens.  This is expected as hydrogenation at $n<$3 does not cause significant distortion of the adsorbed film, hence the vibrational motion of Fe-O-Si bridging oxygen would not be affected. However, at $n=4$ the O$^{\rm ad}$ gets displaced significantly and attaches itself to the Fe-silicate film. This causes a slight shift in the major peak by 13 cm$^{-1}$ to 1015 cm$^{-1}$ due to the strengthening of this bond.

Complete hydrogenation causes a large shift in the smaller peak that is associated with the bending motion of Si-O-Si bridging oxygens. The peak for this mode is around 652 cm$^{-1}$ for Fe$_2$Si$_2$O$_8$H$_2\cdot$O/Ru(0001). \citet{Wodarczyk2013} had observed a Si-O-Si bending mode at 652 cm$^{-1}$, that is the same as ours. However, the Si-O-Si bending mode is observed at 758 cm$^{-1}$ for Fe$_2$Si$_2$O$_8$H$_4\cdot$O/Ru(0001), which indicate a shift of the peak nearly 100 cm$^{-1}$. This indicates a change of character in Si-O tetrahedra which can be due to the differences in the lattice parameters of Fe-silicates \cite{Saritas2021}. The hydrogenation of Fe$_2$Si$_2$O$_8\cdot$O/Ru(0001) also yields additional O-H stretching modes above 3000 cm$^{-1}$. IR stretching modes between 3700-3200 cm$^{-1}$ were observed in various bulk layered silicates \cite{Huang1999}. Since the peaks $>$3000 cm$^{-1}$ are related to hydrogen stretching, they would be completely absent in the $n=0$ hydrogenation. The Fe$_2$Si$_2$O$_8$H$_2\cdot$O/Ru(0001) peak at 3800 cm$^{-1}$ is due to the stretching mode of the H atom bonded to O$^c$ atom, whereas the smaller peak at 3750 cm$^{-1}$ is due to the H bonded to O$^{e, atop}$.  With further hydrogenation, both these peaks shift to lower frequencies. The O$^{e, atop}$ peak shifts rather significantly from 3750 cm$^{-1}$ to 3425 cm$^{-1}$ as shown in Fig. \ref{fgr:rairs}(b). The H-O$^{e, hcp}$ and H-O$^{\rm ad}$ peaks are similar as their convolution creates a larger peak around 3300 cm$^{-1}$. 

\section{Conclusion}
In conclusion, we calculated the structural, energetic, and electronic properties of Fe-silicates on Ru(0001) substrate using density functional theory. It is found that under ambient conditions, the Fe-silicate does hydrogenate with two hydrogens per formula unit, however, further hydrogenation using H$_2$ gas is challenging as it would require very high partial pressures. When complete hydrogenation is achieved, the overlayer structure is potentially exfoliable. We expect that using substrates with weaker oxygen binding energies, the exfoliation energies can be reduced even further. Our subsequent work on Pd substrates will follow. Using partial densities of states and SCLS' energies, we find that the oxidation state of Fe is mostly 3+ throughout different stages of hydrogenation. Simulated RAIRS analysis shows that complete hydrogenation is noticeable in the vibrational spectrum and helps guide the experimental efforts.



\section*{Acknowledgement}
We acknowledge the Army Research Office grant W911NF-19-1-0371 for the funding of this work and also the computational resources provided by the institutional clusters at Yale University.
\bibliography{main}

\begin{thebibliography}{47}%
\makeatletter
\providecommand \@ifxundefined [1]{%
 \@ifx{#1\undefined}
}%
\providecommand \@ifnum [1]{%
 \ifnum #1\expandafter \@firstoftwo
 \else \expandafter \@secondoftwo
 \fi
}%
\providecommand \@ifx [1]{%
 \ifx #1\expandafter \@firstoftwo
 \else \expandafter \@secondoftwo
 \fi
}%
\providecommand \natexlab [1]{#1}%
\providecommand \enquote  [1]{``#1''}%
\providecommand \bibnamefont  [1]{#1}%
\providecommand \bibfnamefont [1]{#1}%
\providecommand \citenamefont [1]{#1}%
\providecommand \href@noop [0]{\@secondoftwo}%
\providecommand \href [0]{\begingroup \@sanitize@url \@href}%
\providecommand \@href[1]{\@@startlink{#1}\@@href}%
\providecommand \@@href[1]{\endgroup#1\@@endlink}%
\providecommand \@sanitize@url [0]{\catcode `\\12\catcode `\$12\catcode
  `\&12\catcode `\#12\catcode `\^12\catcode `\_12\catcode `\%12\relax}%
\providecommand \@@startlink[1]{}%
\providecommand \@@endlink[0]{}%
\providecommand \url  [0]{\begingroup\@sanitize@url \@url }%
\providecommand \@url [1]{\endgroup\@href {#1}{\urlprefix }}%
\providecommand \urlprefix  [0]{URL }%
\providecommand \Eprint [0]{\href }%
\providecommand \doibase [0]{http://dx.doi.org/}%
\providecommand \selectlanguage [0]{\@gobble}%
\providecommand \bibinfo  [0]{\@secondoftwo}%
\providecommand \bibfield  [0]{\@secondoftwo}%
\providecommand \translation [1]{[#1]}%
\providecommand \BibitemOpen [0]{}%
\providecommand \bibitemStop [0]{}%
\providecommand \bibitemNoStop [0]{.\EOS\space}%
\providecommand \EOS [0]{\spacefactor3000\relax}%
\providecommand \BibitemShut  [1]{\csname bibitem#1\endcsname}%
\let\auto@bib@innerbib\@empty
\bibitem [{\citenamefont {B{\"{u}}chner}\ and\ \citenamefont
  {Heyde}(2017)}]{Buchner2017}%
  \BibitemOpen
  \bibfield  {author} {\bibinfo {author} {\bibfnamefont {C.}~\bibnamefont
  {B{\"{u}}chner}}\ and\ \bibinfo {author} {\bibfnamefont {M.}~\bibnamefont
  {Heyde}},\ }\href {\doibase 10.1016/j.progsurf.2017.09.001} {\enquote
  {\bibinfo {title} {{Two-dimensional silica opens new perspectives}},}\ }
  (\bibinfo {year} {2017})\BibitemShut {NoStop}%
\bibitem [{\citenamefont {W{\l}odarczyk}\ \emph {et~al.}(2013)\citenamefont
  {W{\l}odarczyk}, \citenamefont {Sauer}, \citenamefont {Yu}, \citenamefont
  {Boscoboinik}, \citenamefont {Yang}, \citenamefont {Shaikhutdinov},\ and\
  \citenamefont {Freund}}]{Wodarczyk2013}%
  \BibitemOpen
  \bibfield  {author} {\bibinfo {author} {\bibfnamefont {R.}~\bibnamefont
  {W{\l}odarczyk}}, \bibinfo {author} {\bibfnamefont {J.}~\bibnamefont
  {Sauer}}, \bibinfo {author} {\bibfnamefont {X.}~\bibnamefont {Yu}}, \bibinfo
  {author} {\bibfnamefont {J.~A.}\ \bibnamefont {Boscoboinik}}, \bibinfo
  {author} {\bibfnamefont {B.}~\bibnamefont {Yang}}, \bibinfo {author}
  {\bibfnamefont {S.}~\bibnamefont {Shaikhutdinov}}, \ and\ \bibinfo {author}
  {\bibfnamefont {H.~J.}\ \bibnamefont {Freund}},\ }\href {\doibase
  10.1021/ja408772p} {\bibfield  {journal} {\bibinfo  {journal} {J. Am. Chem.
  Soc.}\ }\textbf {\bibinfo {volume} {135}},\ \bibinfo {pages} {19222}
  (\bibinfo {year} {2013})}\BibitemShut {NoStop}%
\bibitem [{\citenamefont {Zhou}\ \emph {et~al.}(2019)\citenamefont {Zhou},
  \citenamefont {Liang}, \citenamefont {Hutchings}, \citenamefont {Fishman},
  \citenamefont {Jhang}, \citenamefont {Li}, \citenamefont {Schwarz},
  \citenamefont {Ismail-Beigi},\ and\ \citenamefont {Altman}}]{Zhou2019}%
  \BibitemOpen
  \bibfield  {author} {\bibinfo {author} {\bibfnamefont {C.}~\bibnamefont
  {Zhou}}, \bibinfo {author} {\bibfnamefont {X.}~\bibnamefont {Liang}},
  \bibinfo {author} {\bibfnamefont {G.~S.}\ \bibnamefont {Hutchings}}, \bibinfo
  {author} {\bibfnamefont {Z.~S.}\ \bibnamefont {Fishman}}, \bibinfo {author}
  {\bibfnamefont {J.-H.}\ \bibnamefont {Jhang}}, \bibinfo {author}
  {\bibfnamefont {M.}~\bibnamefont {Li}}, \bibinfo {author} {\bibfnamefont
  {U.~D.}\ \bibnamefont {Schwarz}}, \bibinfo {author} {\bibfnamefont
  {S.}~\bibnamefont {Ismail-Beigi}}, \ and\ \bibinfo {author} {\bibfnamefont
  {E.~I.}\ \bibnamefont {Altman}},\ }\href {\doibase
  10.1021/acs.chemmater.8b03988} {\bibfield  {journal} {\bibinfo  {journal}
  {Chem. Mater.}\ }\textbf {\bibinfo {volume} {31}},\ \bibinfo {pages} {851}
  (\bibinfo {year} {2019})}\BibitemShut {NoStop}%
\bibitem [{\citenamefont {Fischer}\ \emph {et~al.}(2015)\citenamefont
  {Fischer}, \citenamefont {Sauer}, \citenamefont {Yu}, \citenamefont
  {Boscoboinik}, \citenamefont {Shaikhutdinov},\ and\ \citenamefont
  {Freund}}]{Fischer2015}%
  \BibitemOpen
  \bibfield  {author} {\bibinfo {author} {\bibfnamefont {F.~D.}\ \bibnamefont
  {Fischer}}, \bibinfo {author} {\bibfnamefont {J.}~\bibnamefont {Sauer}},
  \bibinfo {author} {\bibfnamefont {X.}~\bibnamefont {Yu}}, \bibinfo {author}
  {\bibfnamefont {J.~A.}\ \bibnamefont {Boscoboinik}}, \bibinfo {author}
  {\bibfnamefont {S.}~\bibnamefont {Shaikhutdinov}}, \ and\ \bibinfo {author}
  {\bibfnamefont {H.~J.}\ \bibnamefont {Freund}},\ }\href {\doibase
  10.1021/acs.jpcc.5b04291} {\bibfield  {journal} {\bibinfo  {journal} {J.
  Phys. Chem. C}\ }\textbf {\bibinfo {volume} {119}},\ \bibinfo {pages} {15443}
  (\bibinfo {year} {2015})}\BibitemShut {NoStop}%
\bibitem [{\citenamefont {Li}\ \emph {et~al.}(2017)\citenamefont {Li},
  \citenamefont {Tissot}, \citenamefont {Shaikhutdinov},\ and\ \citenamefont
  {Freund}}]{Li2017}%
  \BibitemOpen
  \bibfield  {author} {\bibinfo {author} {\bibfnamefont {L.}~\bibnamefont
  {Li}}, \bibinfo {author} {\bibfnamefont {H.}~\bibnamefont {Tissot}}, \bibinfo
  {author} {\bibfnamefont {S.}~\bibnamefont {Shaikhutdinov}}, \ and\ \bibinfo
  {author} {\bibfnamefont {H.~J.}\ \bibnamefont {Freund}},\ }\href {\doibase
  10.1021/acs.chemmater.6b05213} {\bibfield  {journal} {\bibinfo  {journal}
  {Chem. Mater.}\ }\textbf {\bibinfo {volume} {29}},\ \bibinfo {pages} {931}
  (\bibinfo {year} {2017})}\BibitemShut {NoStop}%
\bibitem [{\citenamefont {Doudin}\ \emph {et~al.}(2021)\citenamefont {Doudin},
  \citenamefont {Saritas}, \citenamefont {Ismail-Beigi},\ and\ \citenamefont
  {Altman}}]{Doudin2021}%
  \BibitemOpen
  \bibfield  {author} {\bibinfo {author} {\bibfnamefont {N.}~\bibnamefont
  {Doudin}}, \bibinfo {author} {\bibfnamefont {K.}~\bibnamefont {Saritas}},
  \bibinfo {author} {\bibfnamefont {S.}~\bibnamefont {Ismail-Beigi}}, \ and\
  \bibinfo {author} {\bibfnamefont {E.~I.}\ \bibnamefont {Altman}},\ }\href
  {\doibase 10.1116/6.0001397} {\bibfield  {journal} {\bibinfo  {journal} {J.
  Vac. Sci. Technol. A}\ }\textbf {\bibinfo {volume} {39}},\ \bibinfo {pages}
  {062201} (\bibinfo {year} {2021})}\BibitemShut {NoStop}%
\bibitem [{\citenamefont {Tissot}\ \emph {et~al.}(2016)\citenamefont {Tissot},
  \citenamefont {Li}, \citenamefont {Shaikhutdinov},\ and\ \citenamefont
  {Freund}}]{Tissot2016}%
  \BibitemOpen
  \bibfield  {author} {\bibinfo {author} {\bibfnamefont {H.}~\bibnamefont
  {Tissot}}, \bibinfo {author} {\bibfnamefont {L.}~\bibnamefont {Li}}, \bibinfo
  {author} {\bibfnamefont {S.}~\bibnamefont {Shaikhutdinov}}, \ and\ \bibinfo
  {author} {\bibfnamefont {H.~J.}\ \bibnamefont {Freund}},\ }\href {\doibase
  10.1039/c6cp03460h} {\bibfield  {journal} {\bibinfo  {journal} {Phys. Chem.
  Chem. Phys.}\ }\textbf {\bibinfo {volume} {18}},\ \bibinfo {pages} {25027}
  (\bibinfo {year} {2016})}\BibitemShut {NoStop}%
\bibitem [{\citenamefont {White}\ \emph {et~al.}(2010)\citenamefont {White},
  \citenamefont {Provis}, \citenamefont {Proffen}, \citenamefont {Riley},\ and\
  \citenamefont {{Van Deventer}}}]{White}%
  \BibitemOpen
  \bibfield  {author} {\bibinfo {author} {\bibfnamefont {C.~E.}\ \bibnamefont
  {White}}, \bibinfo {author} {\bibfnamefont {J.~L.}\ \bibnamefont {Provis}},
  \bibinfo {author} {\bibfnamefont {T.}~\bibnamefont {Proffen}}, \bibinfo
  {author} {\bibfnamefont {D.~P.}\ \bibnamefont {Riley}}, \ and\ \bibinfo
  {author} {\bibfnamefont {J.~S.}\ \bibnamefont {{Van Deventer}}},\ }\href
  {\doibase 10.1021/jp911108d} {\bibfield  {journal} {\bibinfo  {journal} {J.
  Phys. Chem. A}\ }\textbf {\bibinfo {volume} {114}},\ \bibinfo {pages} {4988}
  (\bibinfo {year} {2010})}\BibitemShut {NoStop}%
\bibitem [{\citenamefont {Saritas}\ \emph
  {et~al.}(2021{\natexlab{a}})\citenamefont {Saritas}, \citenamefont {Doudin},
  \citenamefont {Altman},\ and\ \citenamefont {Ismail-Beigi}}]{Saritas2021a}%
  \BibitemOpen
  \bibfield  {author} {\bibinfo {author} {\bibfnamefont {K.}~\bibnamefont
  {Saritas}}, \bibinfo {author} {\bibfnamefont {N.}~\bibnamefont {Doudin}},
  \bibinfo {author} {\bibfnamefont {E.~I.}\ \bibnamefont {Altman}}, \ and\
  \bibinfo {author} {\bibfnamefont {S.}~\bibnamefont {Ismail-Beigi}},\ }\href
  {\doibase 10.1103/PhysRevMaterials.5.104002} {\bibfield  {journal} {\bibinfo
  {journal} {Phys. Rev. Mater.}\ }\textbf {\bibinfo {volume} {5}},\ \bibinfo
  {pages} {104002} (\bibinfo {year} {2021}{\natexlab{a}})},\ \Eprint
  {http://arxiv.org/abs/2011.12938} {arXiv:2011.12938} \BibitemShut {NoStop}%
\bibitem [{\citenamefont {Feibelman}(2002)}]{Feibelman2002}%
  \BibitemOpen
  \bibfield  {author} {\bibinfo {author} {\bibfnamefont {P.~J.}\ \bibnamefont
  {Feibelman}},\ }\href {\doibase 10.1126/science.1065483} {\bibfield
  {journal} {\bibinfo  {journal} {Science (80-. ).}\ }\textbf {\bibinfo
  {volume} {295}},\ \bibinfo {pages} {99} (\bibinfo {year} {2002})}\BibitemShut
  {NoStop}%
\bibitem [{\citenamefont {Michaelides}\ \emph {et~al.}(2003)\citenamefont
  {Michaelides}, \citenamefont {Alavi},\ and\ \citenamefont
  {King}}]{Michaelides2003}%
  \BibitemOpen
  \bibfield  {author} {\bibinfo {author} {\bibfnamefont {A.}~\bibnamefont
  {Michaelides}}, \bibinfo {author} {\bibfnamefont {A.}~\bibnamefont {Alavi}},
  \ and\ \bibinfo {author} {\bibfnamefont {D.~A.}\ \bibnamefont {King}},\
  }\href {\doibase 10.1021/ja028855u} {\bibfield  {journal} {\bibinfo
  {journal} {J. Am. Chem. Soc.}\ }\textbf {\bibinfo {volume} {125}},\ \bibinfo
  {pages} {2746} (\bibinfo {year} {2003})}\BibitemShut {NoStop}%
\bibitem [{\citenamefont {Tatarkhanov}\ \emph {et~al.}(2009)\citenamefont
  {Tatarkhanov}, \citenamefont {Ogletree}, \citenamefont {Rose}, \citenamefont
  {Mitsui}, \citenamefont {Fomin}, \citenamefont {Maier}, \citenamefont {Rose},
  \citenamefont {Cerd{\'{a}}},\ and\ \citenamefont {Salmeron}}]{Tatarkhanov}%
  \BibitemOpen
  \bibfield  {author} {\bibinfo {author} {\bibfnamefont {M.}~\bibnamefont
  {Tatarkhanov}}, \bibinfo {author} {\bibfnamefont {D.~F.}\ \bibnamefont
  {Ogletree}}, \bibinfo {author} {\bibfnamefont {F.}~\bibnamefont {Rose}},
  \bibinfo {author} {\bibfnamefont {T.}~\bibnamefont {Mitsui}}, \bibinfo
  {author} {\bibfnamefont {E.}~\bibnamefont {Fomin}}, \bibinfo {author}
  {\bibfnamefont {S.}~\bibnamefont {Maier}}, \bibinfo {author} {\bibfnamefont
  {M.}~\bibnamefont {Rose}}, \bibinfo {author} {\bibfnamefont {J.~I.}\
  \bibnamefont {Cerd{\'{a}}}}, \ and\ \bibinfo {author} {\bibfnamefont
  {M.}~\bibnamefont {Salmeron}},\ }\href {\doibase 10.1021/ja907468m}
  {\bibfield  {journal} {\bibinfo  {journal} {J. Am. Chem. Soc.}\ }\textbf
  {\bibinfo {volume} {131}},\ \bibinfo {pages} {18425} (\bibinfo {year}
  {2009})}\BibitemShut {NoStop}%
\bibitem [{\citenamefont {Mounet}\ \emph {et~al.}(2018)\citenamefont {Mounet},
  \citenamefont {Gibertini}, \citenamefont {Schwaller}, \citenamefont {Campi},
  \citenamefont {Merkys}, \citenamefont {Marrazzo}, \citenamefont {Sohier},
  \citenamefont {Castelli}, \citenamefont {Cepellotti}, \citenamefont {Pizzi},\
  and\ \citenamefont {Marzari}}]{Mounet2018}%
  \BibitemOpen
  \bibfield  {author} {\bibinfo {author} {\bibfnamefont {N.}~\bibnamefont
  {Mounet}}, \bibinfo {author} {\bibfnamefont {M.}~\bibnamefont {Gibertini}},
  \bibinfo {author} {\bibfnamefont {P.}~\bibnamefont {Schwaller}}, \bibinfo
  {author} {\bibfnamefont {D.}~\bibnamefont {Campi}}, \bibinfo {author}
  {\bibfnamefont {A.}~\bibnamefont {Merkys}}, \bibinfo {author} {\bibfnamefont
  {A.}~\bibnamefont {Marrazzo}}, \bibinfo {author} {\bibfnamefont
  {T.}~\bibnamefont {Sohier}}, \bibinfo {author} {\bibfnamefont {I.~E.}\
  \bibnamefont {Castelli}}, \bibinfo {author} {\bibfnamefont {A.}~\bibnamefont
  {Cepellotti}}, \bibinfo {author} {\bibfnamefont {G.}~\bibnamefont {Pizzi}}, \
  and\ \bibinfo {author} {\bibfnamefont {N.}~\bibnamefont {Marzari}},\ }\href
  {\doibase 10.1038/s41565-017-0035-5} {\bibfield  {journal} {\bibinfo
  {journal} {Nat. Nanotechnol.}\ }\textbf {\bibinfo {volume} {13}},\ \bibinfo
  {pages} {246} (\bibinfo {year} {2018})},\ \Eprint
  {http://arxiv.org/abs/1611.05234} {arXiv:1611.05234} \BibitemShut {NoStop}%
\bibitem [{\citenamefont {Kresse}\ and\ \citenamefont
  {Furthm{\"{u}}ller}(1996{\natexlab{a}})}]{Kresse1996}%
  \BibitemOpen
  \bibfield  {author} {\bibinfo {author} {\bibfnamefont {G.}~\bibnamefont
  {Kresse}}\ and\ \bibinfo {author} {\bibfnamefont {J.}~\bibnamefont
  {Furthm{\"{u}}ller}},\ }\href
  {http://www.sciencedirect.com/science/article/pii/0927025696000080}
  {\bibfield  {journal} {\bibinfo  {journal} {Comput. Mater. Sci.}\ }\textbf
  {\bibinfo {volume} {6}},\ \bibinfo {pages} {15} (\bibinfo {year}
  {1996}{\natexlab{a}})}\BibitemShut {NoStop}%
\bibitem [{\citenamefont {Kresse}\ and\ \citenamefont
  {Furthm{\"{u}}ller}(1996{\natexlab{b}})}]{Kresse1996a}%
  \BibitemOpen
  \bibfield  {author} {\bibinfo {author} {\bibfnamefont {G.}~\bibnamefont
  {Kresse}}\ and\ \bibinfo {author} {\bibfnamefont {J.}~\bibnamefont
  {Furthm{\"{u}}ller}},\ }\href {http://www.ncbi.nlm.nih.gov/pubmed/9984901}
  {\bibfield  {journal} {\bibinfo  {journal} {Phys. Rev. B. Condens. Matter}\
  }\textbf {\bibinfo {volume} {54}},\ \bibinfo {pages} {11169} (\bibinfo {year}
  {1996}{\natexlab{b}})}\BibitemShut {NoStop}%
\bibitem [{\citenamefont {Perdew}\ \emph {et~al.}(1996)\citenamefont {Perdew},
  \citenamefont {Burke},\ and\ \citenamefont {Ernzerhof}}]{Perdew1996}%
  \BibitemOpen
  \bibfield  {author} {\bibinfo {author} {\bibfnamefont {J.~P.}\ \bibnamefont
  {Perdew}}, \bibinfo {author} {\bibfnamefont {K.}~\bibnamefont {Burke}}, \
  and\ \bibinfo {author} {\bibfnamefont {M.}~\bibnamefont {Ernzerhof}},\ }\href
  {http://www.ncbi.nlm.nih.gov/pubmed/10062328} {\bibfield  {journal} {\bibinfo
   {journal} {Phys. Rev. Lett.}\ }\textbf {\bibinfo {volume} {77}},\ \bibinfo
  {pages} {3865} (\bibinfo {year} {1996})}\BibitemShut {NoStop}%
\bibitem [{\citenamefont {Grimme}\ \emph {et~al.}(2010)\citenamefont {Grimme},
  \citenamefont {Antony}, \citenamefont {Ehrlich},\ and\ \citenamefont
  {Krieg}}]{Grimme-d3}%
  \BibitemOpen
  \bibfield  {author} {\bibinfo {author} {\bibfnamefont {S.}~\bibnamefont
  {Grimme}}, \bibinfo {author} {\bibfnamefont {J.}~\bibnamefont {Antony}},
  \bibinfo {author} {\bibfnamefont {S.}~\bibnamefont {Ehrlich}}, \ and\
  \bibinfo {author} {\bibfnamefont {H.}~\bibnamefont {Krieg}},\ }\href
  {\doibase 10.1063/1.3382344} {\bibfield  {journal} {\bibinfo  {journal} {J.
  Chem. Phys.}\ }\textbf {\bibinfo {volume} {132}},\ \bibinfo {pages} {154104}
  (\bibinfo {year} {2010})}\BibitemShut {NoStop}%
\bibitem [{\citenamefont {Kresse}\ and\ \citenamefont
  {Joubert}(1999)}]{Kresse1999}%
  \BibitemOpen
  \bibfield  {author} {\bibinfo {author} {\bibfnamefont {G.}~\bibnamefont
  {Kresse}}\ and\ \bibinfo {author} {\bibfnamefont {D.}~\bibnamefont
  {Joubert}},\ }\href {http://prb.aps.org/abstract/PRB/v59/i3/p1758{\_}1}
  {\bibfield  {journal} {\bibinfo  {journal} {Phys. Rev. B}\ }\textbf {\bibinfo
  {volume} {59}},\ \bibinfo {pages} {11} (\bibinfo {year} {1999})}\BibitemShut
  {NoStop}%
\bibitem [{\citenamefont {Neugebauer}\ and\ \citenamefont
  {Scheffler}(1992)}]{Neugebauer1992a}%
  \BibitemOpen
  \bibfield  {author} {\bibinfo {author} {\bibfnamefont {J.}~\bibnamefont
  {Neugebauer}}\ and\ \bibinfo {author} {\bibfnamefont {M.}~\bibnamefont
  {Scheffler}},\ }\href {\doibase 10.1103/PhysRevB.46.16067} {\bibfield
  {journal} {\bibinfo  {journal} {Phys. Rev. B}\ }\textbf {\bibinfo {volume}
  {46}},\ \bibinfo {pages} {16067} (\bibinfo {year} {1992})}\BibitemShut
  {NoStop}%
\bibitem [{\citenamefont {Karh{\'{a}}nek}\ \emph {et~al.}(2010)\citenamefont
  {Karh{\'{a}}nek}, \citenamefont {Bu{\v{c}}ko},\ and\ \citenamefont
  {Hafner}}]{Karhanek2010}%
  \BibitemOpen
  \bibfield  {author} {\bibinfo {author} {\bibfnamefont {D.}~\bibnamefont
  {Karh{\'{a}}nek}}, \bibinfo {author} {\bibfnamefont {T.}~\bibnamefont
  {Bu{\v{c}}ko}}, \ and\ \bibinfo {author} {\bibfnamefont {J.}~\bibnamefont
  {Hafner}},\ }\href {\doibase 10.1088/0953-8984/22/26/265006} {\bibfield
  {journal} {\bibinfo  {journal} {J. Phys. Condens. Matter}\ }\textbf {\bibinfo
  {volume} {22}},\ \bibinfo {pages} {265006} (\bibinfo {year}
  {2010})}\BibitemShut {NoStop}%
\bibitem [{\citenamefont {L{\"{o}}ffler}\ \emph {et~al.}(2010)\citenamefont
  {L{\"{o}}ffler}, \citenamefont {Uhlrich}, \citenamefont {Baron},
  \citenamefont {Yang}, \citenamefont {Yu}, \citenamefont {Lichtenstein},
  \citenamefont {Heinke}, \citenamefont {B{\"{u}}chner}, \citenamefont {Heyde},
  \citenamefont {Shaikhutdinov}, \citenamefont {Freund}, \citenamefont
  {W{\l}odarczyk}, \citenamefont {Sierka},\ and\ \citenamefont
  {Sauer}}]{Loffler2010}%
  \BibitemOpen
  \bibfield  {author} {\bibinfo {author} {\bibfnamefont {D.}~\bibnamefont
  {L{\"{o}}ffler}}, \bibinfo {author} {\bibfnamefont {J.~J.}\ \bibnamefont
  {Uhlrich}}, \bibinfo {author} {\bibfnamefont {M.}~\bibnamefont {Baron}},
  \bibinfo {author} {\bibfnamefont {B.}~\bibnamefont {Yang}}, \bibinfo {author}
  {\bibfnamefont {X.}~\bibnamefont {Yu}}, \bibinfo {author} {\bibfnamefont
  {L.}~\bibnamefont {Lichtenstein}}, \bibinfo {author} {\bibfnamefont
  {L.}~\bibnamefont {Heinke}}, \bibinfo {author} {\bibfnamefont
  {C.}~\bibnamefont {B{\"{u}}chner}}, \bibinfo {author} {\bibfnamefont
  {M.}~\bibnamefont {Heyde}}, \bibinfo {author} {\bibfnamefont
  {S.}~\bibnamefont {Shaikhutdinov}}, \bibinfo {author} {\bibfnamefont {H.~J.}\
  \bibnamefont {Freund}}, \bibinfo {author} {\bibfnamefont {R.}~\bibnamefont
  {W{\l}odarczyk}}, \bibinfo {author} {\bibfnamefont {M.}~\bibnamefont
  {Sierka}}, \ and\ \bibinfo {author} {\bibfnamefont {J.}~\bibnamefont
  {Sauer}},\ }\href {\doibase 10.1103/PhysRevLett.105.146104} {\bibfield
  {journal} {\bibinfo  {journal} {Phys. Rev. Lett.}\ }\textbf {\bibinfo
  {volume} {105}} (\bibinfo {year} {2010}),\
  10.1103/PhysRevLett.105.146104}\BibitemShut {NoStop}%
\bibitem [{\citenamefont {K{\"{o}}hler}\ and\ \citenamefont
  {Kresse}(2004)}]{Kohler2004}%
  \BibitemOpen
  \bibfield  {author} {\bibinfo {author} {\bibfnamefont {L.}~\bibnamefont
  {K{\"{o}}hler}}\ and\ \bibinfo {author} {\bibfnamefont {G.}~\bibnamefont
  {Kresse}},\ }\href {\doibase 10.1103/PhysRevB.70.165405} {\bibfield
  {journal} {\bibinfo  {journal} {Phys. Rev. B - Condens. Matter Mater. Phys.}\
  }\textbf {\bibinfo {volume} {70}},\ \bibinfo {pages} {1} (\bibinfo {year}
  {2004})}\BibitemShut {NoStop}%
\bibitem [{\citenamefont {Lizzit}\ \emph {et~al.}(2001)\citenamefont {Lizzit},
  \citenamefont {Baraldi}, \citenamefont {Groso}, \citenamefont {Reuter},
  \citenamefont {Ganduglia-Pirovano}, \citenamefont {Stampfl}, \citenamefont
  {Scheffler}, \citenamefont {Stichler}, \citenamefont {Keller}, \citenamefont
  {Wurth},\ and\ \citenamefont {Menzel}}]{Lizzit}%
  \BibitemOpen
  \bibfield  {author} {\bibinfo {author} {\bibfnamefont {S.}~\bibnamefont
  {Lizzit}}, \bibinfo {author} {\bibfnamefont {A.}~\bibnamefont {Baraldi}},
  \bibinfo {author} {\bibfnamefont {A.}~\bibnamefont {Groso}}, \bibinfo
  {author} {\bibfnamefont {K.}~\bibnamefont {Reuter}}, \bibinfo {author}
  {\bibfnamefont {M.~V.}\ \bibnamefont {Ganduglia-Pirovano}}, \bibinfo {author}
  {\bibfnamefont {C.}~\bibnamefont {Stampfl}}, \bibinfo {author} {\bibfnamefont
  {M.}~\bibnamefont {Scheffler}}, \bibinfo {author} {\bibfnamefont
  {M.}~\bibnamefont {Stichler}}, \bibinfo {author} {\bibfnamefont
  {C.}~\bibnamefont {Keller}}, \bibinfo {author} {\bibfnamefont
  {W.}~\bibnamefont {Wurth}}, \ and\ \bibinfo {author} {\bibfnamefont
  {D.}~\bibnamefont {Menzel}},\ }\href {\doibase 10.1103/PhysRevB.63.205419}
  {\bibfield  {journal} {\bibinfo  {journal} {Phys. Rev. B - Condens. Matter
  Mater. Phys.}\ }\textbf {\bibinfo {volume} {63}},\ \bibinfo {pages} {34012}
  (\bibinfo {year} {2001})},\ \Eprint {http://arxiv.org/abs/0102350}
  {arXiv:0102350 [cond-mat]} \BibitemShut {NoStop}%
\bibitem [{\citenamefont {Egelhoff}(1987)}]{Egelhoff1987}%
  \BibitemOpen
  \bibfield  {author} {\bibinfo {author} {\bibfnamefont {W.~F.}\ \bibnamefont
  {Egelhoff}},\ }\href {\doibase 10.1016/0167-5729(87)90007-0} {\enquote
  {\bibinfo {title} {{Core-level binding-energy shifts at surfaces and in
  solids}},}\ } (\bibinfo {year} {1987})\BibitemShut {NoStop}%
\bibitem [{\citenamefont {Jhang}\ \emph {et~al.}(2020)\citenamefont {Jhang},
  \citenamefont {Boscoboinik},\ and\ \citenamefont {Altman}}]{Jhang2020}%
  \BibitemOpen
  \bibfield  {author} {\bibinfo {author} {\bibfnamefont {J.~H.}\ \bibnamefont
  {Jhang}}, \bibinfo {author} {\bibfnamefont {J.~A.}\ \bibnamefont
  {Boscoboinik}}, \ and\ \bibinfo {author} {\bibfnamefont {E.~I.}\ \bibnamefont
  {Altman}},\ }\href {\doibase 10.1063/1.5142621} {\bibfield  {journal}
  {\bibinfo  {journal} {J. Chem. Phys.}\ }\textbf {\bibinfo {volume} {152}},\
  \bibinfo {pages} {084705} (\bibinfo {year} {2020})}\BibitemShut {NoStop}%
\bibitem [{\citenamefont {Wang}\ \emph {et~al.}(2017)\citenamefont {Wang},
  \citenamefont {Zhong}, \citenamefont {Kestell}, \citenamefont {Waluyo},
  \citenamefont {Stacchiola}, \citenamefont {Boscoboinik},\ and\ \citenamefont
  {Lu}}]{Wang2016b}%
  \BibitemOpen
  \bibfield  {author} {\bibinfo {author} {\bibfnamefont {M.}~\bibnamefont
  {Wang}}, \bibinfo {author} {\bibfnamefont {J.~Q.}\ \bibnamefont {Zhong}},
  \bibinfo {author} {\bibfnamefont {J.}~\bibnamefont {Kestell}}, \bibinfo
  {author} {\bibfnamefont {I.}~\bibnamefont {Waluyo}}, \bibinfo {author}
  {\bibfnamefont {D.~J.}\ \bibnamefont {Stacchiola}}, \bibinfo {author}
  {\bibfnamefont {J.~A.}\ \bibnamefont {Boscoboinik}}, \ and\ \bibinfo {author}
  {\bibfnamefont {D.}~\bibnamefont {Lu}},\ }\href {\doibase
  10.1007/s11244-016-0704-x} {\bibfield  {journal} {\bibinfo  {journal} {Top.
  Catal.}\ }\textbf {\bibinfo {volume} {60}},\ \bibinfo {pages} {481} (\bibinfo
  {year} {2017})}\BibitemShut {NoStop}%
\bibitem [{Sup()}]{Supp}%
  \BibitemOpen
  \href@noop {} {}\bibinfo {note} {See Supplementary Information}\BibitemShut
  {NoStop}%
\bibitem [{\citenamefont {Jain}\ \emph {et~al.}(2018)\citenamefont {Jain},
  \citenamefont {Biesinger},\ and\ \citenamefont {Linford}}]{Jain2018}%
  \BibitemOpen
  \bibfield  {author} {\bibinfo {author} {\bibfnamefont {V.}~\bibnamefont
  {Jain}}, \bibinfo {author} {\bibfnamefont {M.~C.}\ \bibnamefont {Biesinger}},
  \ and\ \bibinfo {author} {\bibfnamefont {M.~R.}\ \bibnamefont {Linford}},\
  }\href {\doibase 10.1016/J.APSUSC.2018.03.190} {\bibfield  {journal}
  {\bibinfo  {journal} {Appl. Surf. Sci.}\ }\textbf {\bibinfo {volume} {447}},\
  \bibinfo {pages} {548} (\bibinfo {year} {2018})}\BibitemShut {NoStop}%
\bibitem [{\citenamefont {Baer}\ \emph {et~al.}(2020)\citenamefont {Baer},
  \citenamefont {Artyushkova}, \citenamefont {Cohen}, \citenamefont {Easton},
  \citenamefont {Engelhard}, \citenamefont {Gengenbach}, \citenamefont
  {Greczynski}, \citenamefont {Mack}, \citenamefont {Morgan},\ and\
  \citenamefont {Roberts}}]{Baer2020}%
  \BibitemOpen
  \bibfield  {author} {\bibinfo {author} {\bibfnamefont {D.~R.}\ \bibnamefont
  {Baer}}, \bibinfo {author} {\bibfnamefont {K.}~\bibnamefont {Artyushkova}},
  \bibinfo {author} {\bibfnamefont {H.}~\bibnamefont {Cohen}}, \bibinfo
  {author} {\bibfnamefont {C.~D.}\ \bibnamefont {Easton}}, \bibinfo {author}
  {\bibfnamefont {M.}~\bibnamefont {Engelhard}}, \bibinfo {author}
  {\bibfnamefont {T.~R.}\ \bibnamefont {Gengenbach}}, \bibinfo {author}
  {\bibfnamefont {G.}~\bibnamefont {Greczynski}}, \bibinfo {author}
  {\bibfnamefont {P.}~\bibnamefont {Mack}}, \bibinfo {author} {\bibfnamefont
  {D.~J.}\ \bibnamefont {Morgan}}, \ and\ \bibinfo {author} {\bibfnamefont
  {A.}~\bibnamefont {Roberts}},\ }\href {\doibase 10.1116/6.0000057} {\bibfield
   {journal} {\bibinfo  {journal} {J. Vac. Sci. Technol. A Vacuum, Surfaces,
  Film.}\ }\textbf {\bibinfo {volume} {38}},\ \bibinfo {pages} {031204}
  (\bibinfo {year} {2020})}\BibitemShut {NoStop}%
\bibitem [{\citenamefont {Darwent}\ and\ \citenamefont
  {(U.S.)}(1970)}]{darwent1970bond}%
  \BibitemOpen
  \bibfield  {author} {\bibinfo {author} {\bibfnamefont {B.}~\bibnamefont
  {Darwent}}\ and\ \bibinfo {author} {\bibfnamefont {N.~S. R. D.~S.}\
  \bibnamefont {(U.S.)}},\ }\href
  {https://books.google.com/books?id=4WLJvAEACAAJ} {\emph {\bibinfo {title}
  {Bond Dissociation Energies in Simple Molecules}}},\ NSRDS-NBS\ (\bibinfo
  {publisher} {U.S. Department of Commerce, National Bureau of Standards},\
  \bibinfo {year} {1970})\BibitemShut {NoStop}%
\bibitem [{\citenamefont {{Cox, J.D.; Wagman, D.D.; Medvedev}}(1984)}]{NIST}%
  \BibitemOpen
  \bibfield  {author} {\bibinfo {author} {\bibfnamefont {V.}~\bibnamefont
  {{Cox, J.D.; Wagman, D.D.; Medvedev}}},\ }\href@noop {} {\emph {\bibinfo
  {title} {Hemisph. Publ. Corp., New York}}}\ (\bibinfo {year} {1984})\
  p.~\bibinfo {pages} {1}\BibitemShut {NoStop}%
\bibitem [{\citenamefont {Bolzan}\ \emph {et~al.}(1997)\citenamefont {Bolzan},
  \citenamefont {Fong}, \citenamefont {J},\ and\ \citenamefont
  {Howard}}]{Bolzan1997}%
  \BibitemOpen
  \bibfield  {author} {\bibinfo {author} {\bibfnamefont {A.~A.}\ \bibnamefont
  {Bolzan}}, \bibinfo {author} {\bibfnamefont {C.}~\bibnamefont {Fong}},
  \bibinfo {author} {\bibfnamefont {K.~B.}\ \bibnamefont {J}}, \ and\ \bibinfo
  {author} {\bibfnamefont {C.~J.}\ \bibnamefont {Howard}},\ }\href@noop {}
  {\bibfield  {journal} {\bibinfo  {journal} {Acta Crystallogr.}\ }\textbf
  {\bibinfo {volume} {B53}},\ \bibinfo {pages} {373} (\bibinfo {year}
  {1997})}\BibitemShut {NoStop}%
\bibitem [{\citenamefont {Grimme}(2006)}]{Grimme-d}%
  \BibitemOpen
  \bibfield  {author} {\bibinfo {author} {\bibfnamefont {S.}~\bibnamefont
  {Grimme}},\ }\href {\doibase 10.1002/jcc.20495} {\bibfield  {journal}
  {\bibinfo  {journal} {J. Comput. Chem.}\ }\textbf {\bibinfo {volume} {27}},\
  \bibinfo {pages} {1787} (\bibinfo {year} {2006})}\BibitemShut {NoStop}%
\bibitem [{\citenamefont {Saritas}\ \emph
  {et~al.}(2021{\natexlab{b}})\citenamefont {Saritas}, \citenamefont {Doudin},
  \citenamefont {Altman},\ and\ \citenamefont {Ismail-Beigi}}]{Saritas2021}%
  \BibitemOpen
  \bibfield  {author} {\bibinfo {author} {\bibfnamefont {K.}~\bibnamefont
  {Saritas}}, \bibinfo {author} {\bibfnamefont {N.}~\bibnamefont {Doudin}},
  \bibinfo {author} {\bibfnamefont {E.~I.}\ \bibnamefont {Altman}}, \ and\
  \bibinfo {author} {\bibfnamefont {S.}~\bibnamefont {Ismail-Beigi}},\
  }\href@noop {} {\enquote {\bibinfo {title} {{Stability, Electronic, Magnetic
  and Piezoelectric Properties of Two-dimensional Silicates}},}\ } (\bibinfo
  {year} {2021}{\natexlab{b}})\BibitemShut {NoStop}%
\bibitem [{\citenamefont {Tschersich}\ \emph {et~al.}(2008)\citenamefont
  {Tschersich}, \citenamefont {Fleischhauer},\ and\ \citenamefont
  {Schuler}}]{Tschersich2008}%
  \BibitemOpen
  \bibfield  {author} {\bibinfo {author} {\bibfnamefont {K.~G.}\ \bibnamefont
  {Tschersich}}, \bibinfo {author} {\bibfnamefont {J.~P.}\ \bibnamefont
  {Fleischhauer}}, \ and\ \bibinfo {author} {\bibfnamefont {H.}~\bibnamefont
  {Schuler}},\ }\href {\doibase 10.1063/1.2963956} {\bibfield  {journal}
  {\bibinfo  {journal} {J. Appl. Phys.}\ }\textbf {\bibinfo {volume} {104}},\
  \bibinfo {pages} {034908} (\bibinfo {year} {2008})}\BibitemShut {NoStop}%
\bibitem [{\citenamefont {Gibertini}\ \emph {et~al.}(2019)\citenamefont
  {Gibertini}, \citenamefont {Koperski}, \citenamefont {Morpurgo},\ and\
  \citenamefont {Novoselov}}]{Gibertini2019}%
  \BibitemOpen
  \bibfield  {author} {\bibinfo {author} {\bibfnamefont {M.}~\bibnamefont
  {Gibertini}}, \bibinfo {author} {\bibfnamefont {M.}~\bibnamefont {Koperski}},
  \bibinfo {author} {\bibfnamefont {A.~F.}\ \bibnamefont {Morpurgo}}, \ and\
  \bibinfo {author} {\bibfnamefont {K.~S.}\ \bibnamefont {Novoselov}},\ }\href
  {\doibase 10.1038/s41565-019-0438-6} {\bibfield  {journal} {\bibinfo
  {journal} {Nat. Nanotechnol.}\ }\textbf {\bibinfo {volume} {14}},\ \bibinfo
  {pages} {408} (\bibinfo {year} {2019})}\BibitemShut {NoStop}%
\bibitem [{\citenamefont {B{\"{o}}ttcher}\ and\ \citenamefont
  {Niehus}(1999)}]{Bo}%
  \BibitemOpen
  \bibfield  {author} {\bibinfo {author} {\bibfnamefont {A.}~\bibnamefont
  {B{\"{o}}ttcher}}\ and\ \bibinfo {author} {\bibfnamefont {H.}~\bibnamefont
  {Niehus}},\ }\href {\doibase 10.1103/PhysRevB.60.14396} {\bibfield  {journal}
  {\bibinfo  {journal} {Phys. Rev. B - Condens. Matter Mater. Phys.}\ }\textbf
  {\bibinfo {volume} {60}},\ \bibinfo {pages} {14396} (\bibinfo {year}
  {1999})}\BibitemShut {NoStop}%
\bibitem [{\citenamefont {Himpsel}\ \emph {et~al.}(1982)\citenamefont
  {Himpsel}, \citenamefont {Christmann}, \citenamefont {Heimann}, \citenamefont
  {Eastman},\ and\ \citenamefont {Feibelman}}]{Himpsel1982}%
  \BibitemOpen
  \bibfield  {author} {\bibinfo {author} {\bibfnamefont {F.}~\bibnamefont
  {Himpsel}}, \bibinfo {author} {\bibfnamefont {K.}~\bibnamefont {Christmann}},
  \bibinfo {author} {\bibfnamefont {P.}~\bibnamefont {Heimann}}, \bibinfo
  {author} {\bibfnamefont {D.}~\bibnamefont {Eastman}}, \ and\ \bibinfo
  {author} {\bibfnamefont {P.~J.}\ \bibnamefont {Feibelman}},\ }\href {\doibase
  10.1016/0167-2584(82)90679-X} {\bibfield  {journal} {\bibinfo  {journal}
  {Surf. Sci. Lett.}\ }\textbf {\bibinfo {volume} {115}},\ \bibinfo {pages}
  {L159} (\bibinfo {year} {1982})}\BibitemShut {NoStop}%
\bibitem [{\citenamefont {Maintz}\ \emph {et~al.}(2016)\citenamefont {Maintz},
  \citenamefont {Deringer}, \citenamefont {Tchougr{\'{e}}eff},\ and\
  \citenamefont {Dronskowski}}]{Maintz2016}%
  \BibitemOpen
  \bibfield  {author} {\bibinfo {author} {\bibfnamefont {S.}~\bibnamefont
  {Maintz}}, \bibinfo {author} {\bibfnamefont {V.~L.}\ \bibnamefont
  {Deringer}}, \bibinfo {author} {\bibfnamefont {A.~L.}\ \bibnamefont
  {Tchougr{\'{e}}eff}}, \ and\ \bibinfo {author} {\bibfnamefont
  {R.}~\bibnamefont {Dronskowski}},\ }\href {\doibase 10.1002/jcc.24300}
  {\bibfield  {journal} {\bibinfo  {journal} {J. Comput. Chem.}\ }\textbf
  {\bibinfo {volume} {37}},\ \bibinfo {pages} {1030} (\bibinfo {year}
  {2016})}\BibitemShut {NoStop}%
\bibitem [{\citenamefont {Dronskowski}\ and\ \citenamefont
  {Bloechl}(1993)}]{Dronskowski1993}%
  \BibitemOpen
  \bibfield  {author} {\bibinfo {author} {\bibfnamefont {R.}~\bibnamefont
  {Dronskowski}}\ and\ \bibinfo {author} {\bibfnamefont {P.~E.}\ \bibnamefont
  {Bloechl}},\ }\href {\doibase 10.1021/j100135a014} {\bibfield  {journal}
  {\bibinfo  {journal} {J. Phys. Chem.}\ }\textbf {\bibinfo {volume} {97}},\
  \bibinfo {pages} {8617} (\bibinfo {year} {1993})}\BibitemShut {NoStop}%
\bibitem [{\citenamefont {Maintz}\ \emph {et~al.}(2013)\citenamefont {Maintz},
  \citenamefont {Deringer}, \citenamefont {Tchougr{\'{e}}eff},\ and\
  \citenamefont {Dronskowski}}]{Maintz2013}%
  \BibitemOpen
  \bibfield  {author} {\bibinfo {author} {\bibfnamefont {S.}~\bibnamefont
  {Maintz}}, \bibinfo {author} {\bibfnamefont {V.~L.}\ \bibnamefont
  {Deringer}}, \bibinfo {author} {\bibfnamefont {A.~L.}\ \bibnamefont
  {Tchougr{\'{e}}eff}}, \ and\ \bibinfo {author} {\bibfnamefont
  {R.}~\bibnamefont {Dronskowski}},\ }\href {\doibase 10.1002/jcc.23424}
  {\bibfield  {journal} {\bibinfo  {journal} {J. Comput. Chem.}\ }\textbf
  {\bibinfo {volume} {34}},\ \bibinfo {pages} {2557} (\bibinfo {year}
  {2013})}\BibitemShut {NoStop}%
\bibitem [{\citenamefont {Raebiger}\ \emph {et~al.}(2008)\citenamefont
  {Raebiger}, \citenamefont {Lany},\ and\ \citenamefont
  {Zunger}}]{Raebiger2008}%
  \BibitemOpen
  \bibfield  {author} {\bibinfo {author} {\bibfnamefont {H.}~\bibnamefont
  {Raebiger}}, \bibinfo {author} {\bibfnamefont {S.}~\bibnamefont {Lany}}, \
  and\ \bibinfo {author} {\bibfnamefont {A.}~\bibnamefont {Zunger}},\ }\href
  {\doibase 10.1038/nature07009} {\bibfield  {journal} {\bibinfo  {journal}
  {Nature}\ }\textbf {\bibinfo {volume} {453}},\ \bibinfo {pages} {763}
  (\bibinfo {year} {2008})}\BibitemShut {NoStop}%
\bibitem [{\citenamefont {Fongkaew}\ \emph {et~al.}(2017)\citenamefont
  {Fongkaew}, \citenamefont {Akrobetu}, \citenamefont {Sehirlioglu},
  \citenamefont {Voevodin}, \citenamefont {Limpijumnong},\ and\ \citenamefont
  {Lambrecht}}]{Fongkaew2017}%
  \BibitemOpen
  \bibfield  {author} {\bibinfo {author} {\bibfnamefont {I.}~\bibnamefont
  {Fongkaew}}, \bibinfo {author} {\bibfnamefont {R.}~\bibnamefont {Akrobetu}},
  \bibinfo {author} {\bibfnamefont {A.}~\bibnamefont {Sehirlioglu}}, \bibinfo
  {author} {\bibfnamefont {A.}~\bibnamefont {Voevodin}}, \bibinfo {author}
  {\bibfnamefont {S.}~\bibnamefont {Limpijumnong}}, \ and\ \bibinfo {author}
  {\bibfnamefont {W.~R.}\ \bibnamefont {Lambrecht}},\ }\href {\doibase
  10.1016/J.ELSPEC.2017.05.009} {\bibfield  {journal} {\bibinfo  {journal} {J.
  Electron Spectros. Relat. Phenomena}\ }\textbf {\bibinfo {volume} {218}},\
  \bibinfo {pages} {21} (\bibinfo {year} {2017})}\BibitemShut {NoStop}%
\bibitem [{\citenamefont {Pasquarello}\ \emph {et~al.}(1995)\citenamefont
  {Pasquarello}, \citenamefont {Hybertsen},\ and\ \citenamefont
  {Car}}]{Pasquarello1995}%
  \BibitemOpen
  \bibfield  {author} {\bibinfo {author} {\bibfnamefont {A.}~\bibnamefont
  {Pasquarello}}, \bibinfo {author} {\bibfnamefont {M.~S.}\ \bibnamefont
  {Hybertsen}}, \ and\ \bibinfo {author} {\bibfnamefont {R.}~\bibnamefont
  {Car}},\ }\href@noop {} {\bibfield  {journal} {\bibinfo  {journal} {Phys.
  Rev. Lett.}\ }\textbf {\bibinfo {volume} {74}},\ \bibinfo {pages} {1024}
  (\bibinfo {year} {1995})}\BibitemShut {NoStop}%
\bibitem [{\citenamefont {Wang}\ \emph {et~al.}(2019)\citenamefont {Wang},
  \citenamefont {Du}, \citenamefont {Sushko}, \citenamefont {Bowden},
  \citenamefont {Stoerzinger}, \citenamefont {Heald}, \citenamefont {Scafetta},
  \citenamefont {Kaspar},\ and\ \citenamefont {Chambers}}]{Wang2019}%
  \BibitemOpen
  \bibfield  {author} {\bibinfo {author} {\bibfnamefont {L.}~\bibnamefont
  {Wang}}, \bibinfo {author} {\bibfnamefont {Y.}~\bibnamefont {Du}}, \bibinfo
  {author} {\bibfnamefont {P.~V.}\ \bibnamefont {Sushko}}, \bibinfo {author}
  {\bibfnamefont {M.~E.}\ \bibnamefont {Bowden}}, \bibinfo {author}
  {\bibfnamefont {K.~A.}\ \bibnamefont {Stoerzinger}}, \bibinfo {author}
  {\bibfnamefont {S.~M.}\ \bibnamefont {Heald}}, \bibinfo {author}
  {\bibfnamefont {M.~D.}\ \bibnamefont {Scafetta}}, \bibinfo {author}
  {\bibfnamefont {T.~C.}\ \bibnamefont {Kaspar}}, \ and\ \bibinfo {author}
  {\bibfnamefont {S.~A.}\ \bibnamefont {Chambers}},\ }\href {\doibase
  10.1103/PhysRevMaterials.3.025401} {\bibfield  {journal} {\bibinfo  {journal}
  {Phys. Rev. Materials}\ }\textbf {\bibinfo {volume} {3}},\ \bibinfo {pages}
  {025401} (\bibinfo {year} {2019})}\BibitemShut {NoStop}%
\bibitem [{\citenamefont {Yang}\ \emph {et~al.}(2012)\citenamefont {Yang},
  \citenamefont {Kaden}, \citenamefont {Yu}, \citenamefont {Boscoboinik},
  \citenamefont {Martynova}, \citenamefont {Lichtenstein}, \citenamefont
  {Heyde}, \citenamefont {Sterrer}, \citenamefont {W{\l}odarczyk},
  \citenamefont {Sierka}, \citenamefont {Sauer}, \citenamefont
  {Shaikhutdinov},\ and\ \citenamefont {Freund}}]{Yang2012}%
  \BibitemOpen
  \bibfield  {author} {\bibinfo {author} {\bibfnamefont {B.}~\bibnamefont
  {Yang}}, \bibinfo {author} {\bibfnamefont {W.~E.}\ \bibnamefont {Kaden}},
  \bibinfo {author} {\bibfnamefont {X.}~\bibnamefont {Yu}}, \bibinfo {author}
  {\bibfnamefont {J.~A.}\ \bibnamefont {Boscoboinik}}, \bibinfo {author}
  {\bibfnamefont {Y.}~\bibnamefont {Martynova}}, \bibinfo {author}
  {\bibfnamefont {L.}~\bibnamefont {Lichtenstein}}, \bibinfo {author}
  {\bibfnamefont {M.}~\bibnamefont {Heyde}}, \bibinfo {author} {\bibfnamefont
  {M.}~\bibnamefont {Sterrer}}, \bibinfo {author} {\bibfnamefont
  {R.}~\bibnamefont {W{\l}odarczyk}}, \bibinfo {author} {\bibfnamefont
  {M.}~\bibnamefont {Sierka}}, \bibinfo {author} {\bibfnamefont
  {J.}~\bibnamefont {Sauer}}, \bibinfo {author} {\bibfnamefont
  {S.}~\bibnamefont {Shaikhutdinov}}, \ and\ \bibinfo {author} {\bibfnamefont
  {H.~J.}\ \bibnamefont {Freund}},\ }\href {\doibase 10.1039/C2CP41355H}
  {\bibfield  {journal} {\bibinfo  {journal} {Phys. Chem. Chem. Phys.}\
  }\textbf {\bibinfo {volume} {14}},\ \bibinfo {pages} {11344} (\bibinfo {year}
  {2012})}\BibitemShut {NoStop}%
\bibitem [{\citenamefont {Huang}(1999)}]{Huang1999}%
  \BibitemOpen
  \bibfield  {author} {\bibinfo {author} {\bibfnamefont {Y.}~\bibnamefont
  {Huang}},\ }\href {\doibase 10.1021/cm980403m} {\bibfield  {journal}
  {\bibinfo  {journal} {Chem. Mater.}\ }\textbf {\bibinfo {volume} {11}},\
  \bibinfo {pages} {1210} (\bibinfo {year} {1999})}\BibitemShut {NoStop}%
\end{thebibliography}%

\end{document}